\documentclass[10pt,journal]{IEEEtran}
\usepackage{outline}
\usepackage{pmgraph}
\usepackage[normalem]{ulem}
\usepackage[utf8]{inputenc}
\usepackage{amssymb}
\usepackage{hyperref}
\usepackage{amsmath}
\usepackage{graphicx}
\usepackage{times}
\usepackage{xcolor}
\usepackage{xspace}
\usepackage[colorinlistoftodos]{todonotes} % for \todo
\usepackage{cite}
\usepackage{bm} % bold
\usepackage{url}

\usepackage{epstopdf}
\AppendGraphicsExtensions{.tiff}
\graphicspath{{fig/}} % Directory in which figures are stored

\usepackage{epsfig}
\usepackage{tikz}
\usetikzlibrary{spy}
\usepackage{algpseudocode}
\usepackage{algorithm}
\usepackage{mathrsfs}

\usepackage{multirow}%,setspace,verbatim,amsfonts,graphicx,amsmath,amsthm,amsbsy,amssymb,epsfig,url,cite}
\def\QED{~\rule[-1pt]{5pt}{5pt}\par\medskip}

\long\def\comment#1{} % comment out text

\newcommand{\xmath}[1] {\ensuremath{#1}\xspace}
\newcommand{\blmath}[1] {\xmath{\bm{#1}}}

\newcommand{\x}{\blmath{x}}

\newcommand{\ab}{{\blmath a}}

\newcommand{\db}{{\blmath d}}
\newcommand{\eb}{{\blmath e}}

\newcommand{\Lc}{\mathcal{L}}
\newcommand{\Tc}{\mathcal{T}}
\newcommand{\Xc}{\mathcal{X}}
\newcommand{\Yc}{\mathcal{Y}}

\newcommand{\Thetab}{{\boldsymbol {\Theta}}}

\newcommand{\Rd}{{\mathbb R}}

\newcommand{\thetab}{{\boldsymbol {\theta}}}

\newcommand{\Pc}{{{\mathcal P}}}

\newcommand{\beq}{\begin{equation}}
\newcommand{\eeq}{\end{equation}}
\newcommand{\beqa}{\begin{eqnarray}}
\newcommand{\eeqa}{\end{eqnarray}}
\newcommand{\Fc}{{\mathcal F}}

\newcommand{\Sd}{{\mathbb S}}

\newcommand{\omegab}{\boldsymbol{\omega}}

\DeclareMathOperator{\sgn}{sgn}
\newcommand{\degree}{^\circ}

\newcommand{\added}[1] {\textcolor{black}{#1}} % for revision

\begin{document}

\title{Differentiated Backprojection Domain Deep Learning for Conebeam Artifact Removal} %  using Differentiated Backprojection }
%\date{\vspace{-4ex}}

\author{Yoseob~Han,~%\IEEEmembership{Student Member,~IEEE,}
        Junyoung Kim,~%\IEEEmembership{Member,~IEEE,}
        and~Jong~Chul~Ye,~\IEEEmembership{Fellow,~IEEE}% <-this % stops a space
\thanks{Y. Han is with the Theoretical Division, Los Alamos National Laboratory, Los Alamos, NM 87545, USA
		(e-mail: yoseobhan@lanl.gov).
J. Kim, and J. C. Ye are with the Department of Bio and Brain Engineering, Korea Advanced Institute of Science and Technology (KAIST), 
		Daejeon 34141, Republic of Korea 
		(e-mail: \{junyoung.kim, jong.ye\}@kaist.ac.kr). 
		J.C. Ye is also with the Department of Mathematical Sciences, KAIST.
		This work was performed when the first author was at KAIST.
		The first two authors contributed equally to this work.
		
		This work was supported by the Technology development Program (1425117734) funded by the Ministry of SMEs and Startups (MSS, Korea).
		
		Copyright (c) 2019 IEEE. Personal use of this material is permitted. However, permission to use this material for any other purposes must be obtained from the IEEE by sending a request to pubs-permissions@ieee.org.}%
	}

\maketitle
%\IEEEpubid{\begin{minipage}{\textwidth}\ \\[40pt]
%Copyright (c) 2019 IEEE. Personal use of this material is permitted. However, permission to use this material for any other purposes must be obtained from the IEEE by sending a request to pubs-permissions@ieee.org.
%\end{minipage}}

% As a general rule, do not put math, special symbols or citations in the abstract or keywords.
\begin{abstract}
Conebeam CT using a circular trajectory  is quite often used for various applications due to its relative simple geometry.
For conebeam geometry,
Feldkamp, Davis and Kress  algorithm is regarded as the standard reconstruction method, but this algorithm suffers from
so-called conebeam artifacts as the cone angle increases. Various model-based iterative reconstruction methods have been developed to reduce the cone-beam artifacts, but these algorithms usually
 require multiple  applications of computational expensive  forward and backprojections. % in spite of generally not satisfactory reconstruction results.
In this paper, we develop a novel deep learning approach for accurate conebeam artifact removal.
In particular,  our deep network,  designed on the differentiated backprojection  domain,  performs a data-driven  inversion of an ill-posed deconvolution
 problem associated
with the
Hilbert transform.
The reconstruction results  along the coronal and sagittal directions are then combined using a spectral blending technique to minimize the spectral leakage.
 %Numerical 
 Experimental results under various conditions confirmed that our method generalizes well
 and outperforms the existing iterative methods despite significantly reduced runtime complexity.

\end{abstract}


\IEEEpeerreviewmaketitle

\section{Introduction}\label{sec:introduction}
\IEEEPARstart{C}{onebeam} X-ray CT with a large number of detector rows is often used for  interventional imaging, dental CT, etc,
due to the capability of obtaining high-resolution projection images with a relative simple scanner geometry.
For the conebeam geometry, Feldkamp, Davis and Kress (FDK) algorithm  \cite{feldkamp1984practical} has been extensively used as a standard reconstruction method.
%However, aside from the increased scatters due to the lack of the detector collimators, the conebeam CT with
Unfortunately, the FDK algorithm for  conebeam CT
usually suffers from conebeam artifacts as the cone-angle increases.
To address this problem, some researchers have proposed modified FDK algorithms by introducing
angle-dependent weighting on the measured projection data \cite{grass2001angular, tang2005three, mori2006combination }. 
However,  these methods usually  work only for small cone angles.

Mathematically, conebeam artifacts arise from  inherent defects in the circular trajectory that does not satisfy the Tuy's condition \cite{tuy1983inversion}.
In Fourier domain, this is manifested as the missing spectral components at the specific frequencies
that are determined by the scanner geometry \cite{bartolac2009local,pack2013mitigating,peyrin1992analysis}.
Accordingly, without the use of additional prior information, accurate removal of the conebeam artifacts may not be feasible.

To address this issue,
several  model-based iterative reconstruction (MBIR) methods \cite{sidky2008image, sidky2012convex, zhang2016artifact, xia2016optimization} have been proposed by imposing total variation (TV) and other penalties. The role of the penalty function is to impose the constraint to compensate for the missing frequency. Unfortunately,  these  algorithms are computationally expensive due to the iterative applications of 3-D projection and backprojection.

In recent years, deep learning approaches have been successfully used for a variety of applications from image classification to  medical image reconstruction \cite{krizhevsky2012imagenet,ronneberger2015u,han2016deep,kang2017deep,chen2017low,kang2018deep,jin2017deep,han2018framing,han2018roi,han2018one}. In X-ray CT, various deep learning reconstruction methods have been developed   for low-dose CT \cite{kang2017deep,chen2017low,kang2018deep}, sparse-view CT \cite{han2016deep,jin2017deep,han2018framing}, interior tomography \cite{han2018roi,han2018one}, and so on. 
%These deep learning approaches usually outperform MBIR methods in their image quality and  reconstruction time. 
In addition, in recent theoretical works \cite{ye2018deep, ye2019cnn}, the authors showed that a deep convolutional neural network with an encoder-decoder structure  is related to
 a novel frame expansion using combinatorial convolutional frames.
More specifically, thanks to the rectified linear unit (ReLU) nonlinearities, the input space is 
partitioned into large number of non-overlapping regions so that input images for each region share the same linear frame representation  \cite{ye2019cnn}. Therefore, once a neural network is trained,  each input image can automatically choose the appropriate linear
representation.
%in a real time manner.

\begin{figure*}[!h] 	
\centerline{\includegraphics[width=1\linewidth]{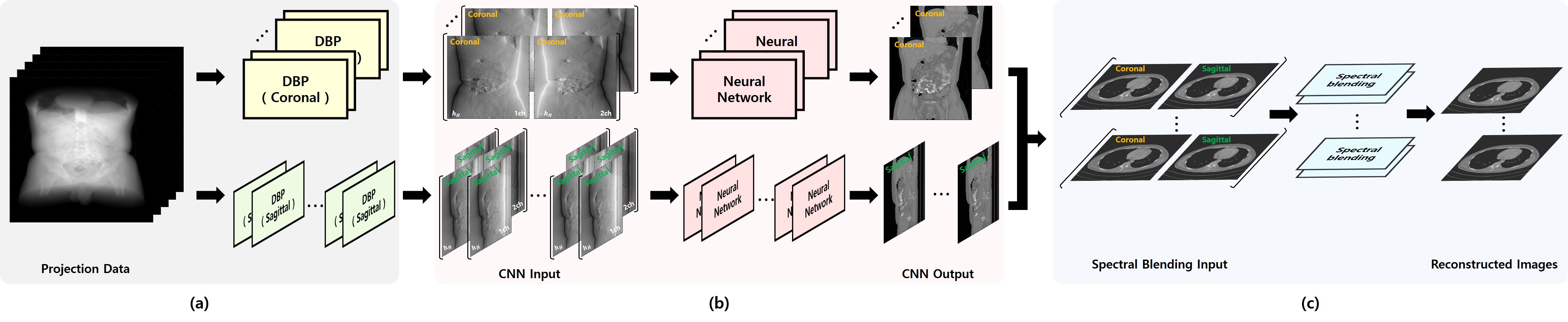}}
\caption{Overview for differentiated backrpoejction domain deep learning. (a) shows the differentiated backprojection parts of coronal and sagittal directions, respectively. (b) illustrates the neural network part, and (c) shows the spectral blending.}
\vspace*{-0.5cm}
\label{fig:overview}
\end{figure*}

Inspired by these findings,  here we propose a deep neural network for conebeam artifact removal.
Instead of using standard image domain implementation of deep neural networks, we are aiming for a further step.
Specifically, we have shown in our prior work \cite{lee2015interior}  that
the conebeam reconstruction problem can be solved by successive 2-D reconstruction in the differentiated backprojection (DBP) domain and subsequent spectral mixing to minimize spectral loss.
This was due to the the exact factorization of the initial 3-D conebeam reconstruction problem into
a set of independent 2-D inversion problems as proposed by \cite{dennerlein2008factorization}.
Furthermore,  in our recent paper \cite{han2018one} for  interior tomography problem, we have also demonstrated that deep neural networks on the DBP domain generalize better than the image domain approaches   \cite{han2018roi}.
By synergistically combining these findings, we propose a DBP domain deep learning framework for cone-beam artifact removal. 
Specifically,  as shown in Fig.~\ref{fig:overview}, the input of our neural networks is the DBP images either in the {coronal and sagittal} directions, and our  deep neural
network is trained to solve a deconvolution problem involving Hilbert transform. Finally, the reconstructed  3-D volumes along the coronal and sagittal directional
DBP are blended into a one 3-D volume
using spectral mixing techniques. % illustrates the overall reconstruction flow.   % in axial domain.

Similar to our prior observation from ROI tomography \cite{han2018roi}, this paper shows that
the neural network trained on the DBP domain generalizes better and can remove  conebeam artifact accurately
under various conditions that are not considered during the training phase.
For example, a neural network trained on a specific  cone-angle can be used for different cone-angles without retraining the neural network,
and the neural network trained on the noiseless simulation data can be successfully used for noisy  real measurements.
The origin of such improved generalization
will be  discussed. % in term of inductive bias originated from  piecewise linearity of the neural network}. %, which results in inherent regularization effects. }

This paper is structured as follows. In Section \ref{sec:math}, the  mathematical preliminaries for  conebeam CT using a circular trajectory
are provided.
Section~\ref{sec:theory} then discusses our main contributions: the DBP domain deep neural network and spectral blending.
The experimental data set and the neural network training are explained in Section~\ref{sec:method}.  The experimental results are given
 in Section \ref{sec:result}, which is followed by discussion and conclusions  in Sections  \ref{sec:discussion} and \ref{sec:conclusion}, respectively.

\section{Mathematical Preliminaries}\label{sec:math}

\subsection{Notation}
Let $\x=(x,y,z)$ denote the point in $\Rd^3$, where $x,y$ and $z$ are
the Cartesian coordinates  with respect 
to the patient. 
For the circular trajectory,  the X-ray source
rotates around the  object $f(\x)$ such that the  source trajectory
is represented by
\begin{eqnarray}
\ab(\lambda) = \begin{bmatrix} R\cos\lambda & R\sin\lambda & 0 \end{bmatrix}^\top,\quad \lambda\in [0,2\pi],
\end{eqnarray}
where $R$ denotes the radius of the circular scan.
In this paper, the symbol  $^\top$ denotes
the transpose of a matrix or vector.

The X-ray transform $D_f$, which maps $f(\x)$ into the set of its line integrals, is defined as
%is defined by
\begin{equation}\label{eq:D_f}
D_f(\ab,\thetab)=\int_{-\infty}^\infty dt~f(\ab+t\thetab)~,
\end{equation}
where
$\thetab$ denote a vector on the unit sphere $\Sd\in \Rd^3$.
The 3-D inverse Fourier transform is defined by
\begin{equation}\label{eq:inv}
f(\x)=\frac{1}{(2\pi)^3}\int d\omegab~\hat f(\omegab)e^{\iota\omegab^\top\x},
\end{equation}
where $\hat f(\omegab)$ denotes the Fourier transform of $f(\x)$,  $\omegab\in \Rd^3$ refers to the angular frequency,
 and $\iota=\sqrt{-1}$.

\subsection{Differentiated Backprojection (DBP)}

For a given X-ray source trajectory $\ab(\lambda)$,  the differentiated backprojection (DBP) on a point $\x\in \Rd^3$  using the projection data from 
x-ray source trajectory $\ab(\lambda),\lambda \in [\lambda^-,\lambda^+]$ is %specified by $\lambda^-:=\lambda^-(\x)$ and $\lambda^+:=\lambda^+(\x)$$ is 
 defined as \cite{pack2005cone,zou2004exact,zou2004image}:
\begin{equation}\label{eq:g}
g(\x)=\int_{\lambda^-}^{\lambda^+}d\lambda~\frac{1}{\|\x-\ab(\lambda)\|}\left.
\frac{\partial}{\partial\mu}D_f(\ab(\mu),\thetab)\right|_{\mu=\lambda},
\end{equation}
where %$\x$ is on the chord line %,$[\lambda^-,\lambda^+]\subset[\lambda_{\min},\lambda_{\max}]$  is an interval from the source trajectories, 
%and 
$1/\|\x-\ab(\lambda)\|$ is the magnification factor dependent weighting. 
One of the most fundamental properties of the DBP is its relationship to Fourier transform \eqref{eq:inv}. 
Specifically,  %since it
the DBP data in \eqref{eq:g} 
can be represented as \cite{lee2015interior,zou2004image}:
\begin{equation}\label{eq:gfourier}
g(\x)=\frac{1}{(2\pi)^3}\int d\omegab~\hat f(\omegab)e^{\iota\omegab^\top\x}\sigma(\x,\omegab,\lambda^\pm),
\end{equation}
where
\begin{eqnarray}
{\sigma(\x,\omegab,\lambda^\pm)=\iota\pi[\sgn(\omegab^\top\db^{-}(\x))-\sgn(\omegab^\top\db^{+}(\x))], }
\label{eq:sigma}
\end{eqnarray}
and
\begin{eqnarray}
{\db^{+}(\x) =\frac{\x-\ab(\lambda^+)}{\|\x-\ab(\lambda^+)\|},\quad \db^{-}(\x) =\frac{\x-\ab(\lambda^-)}{\|\x-\ab(\lambda^-)\|} }
\end{eqnarray}
%To make this paper self-contained, the proof of \eqref{eq:gfourier} is provided in Appendix.
Compared to the Fourier formulation in \eqref{eq:inv}, the main difference of \eqref{eq:gfourier} is the additional term $\sigma(\x,\omegab,\lambda^\pm)$.
This suggests that DBP provides a filtered version of $f(\x)$.
Since the spectrum of the filter
$\sigma(\x,\omegab,\lambda^\pm)$ also depends on $\x$, the corresponding filter is a spatially varying filter.

\begin{figure}[!b] 	
\centerline{\includegraphics[width=0.9\linewidth]{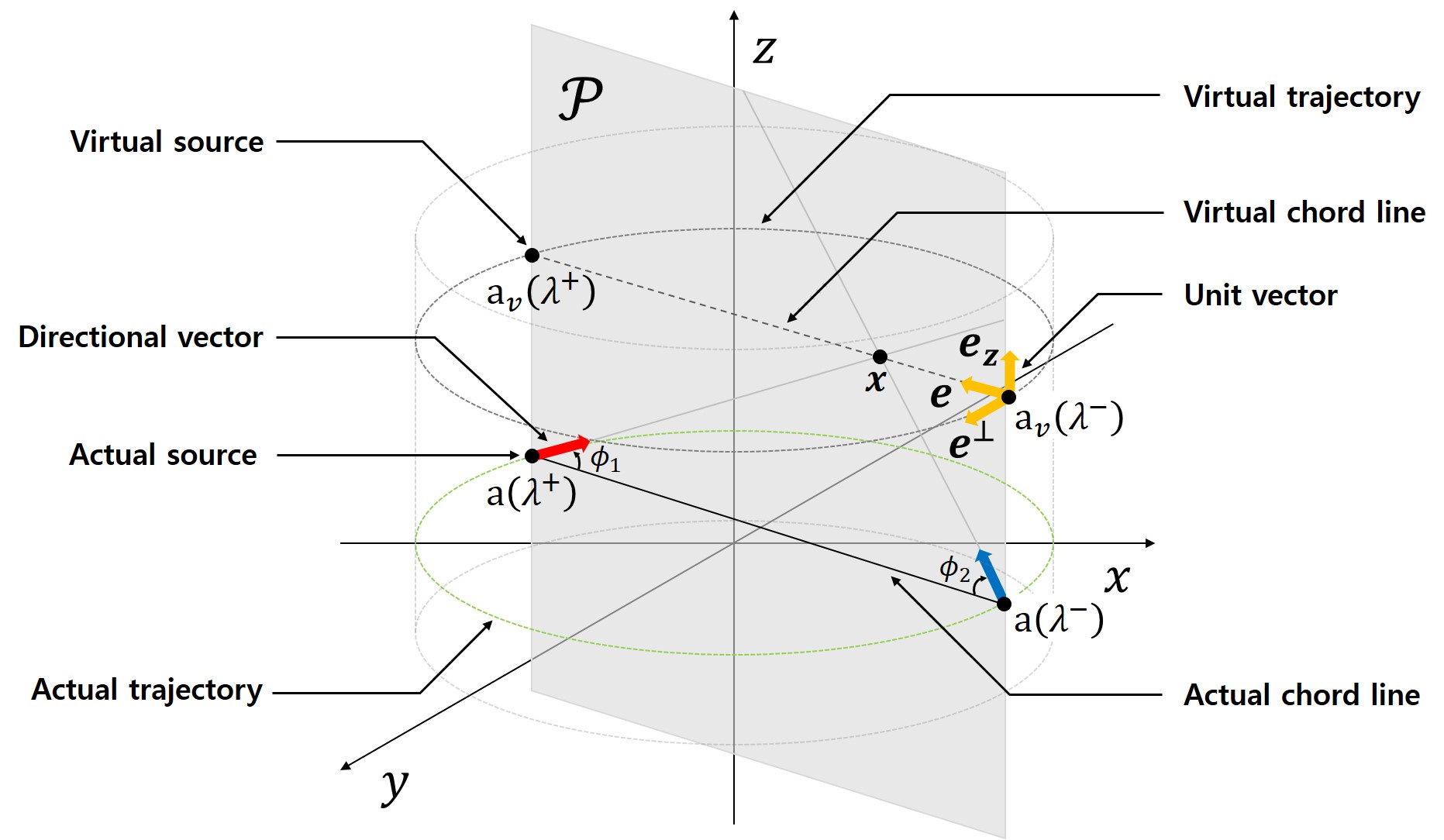}}
\caption{Coordinate system  on a factorized conebeam geometry. }
\vspace*{-0.5cm}
\label{fig:trajectory}
\end{figure}

\subsection{Factorized Representation of Conebeam Geometry}

Inspired by the relationship in \eqref{eq:gfourier},
Dennerlein et al \cite{dennerlein2008factorization} proposed a factorization
method for the conebeam CT with the circular trajectory.  Similar idea was proposed by Yu et al \cite{yu2006region}, so we introduce the key ideas of
these factorization methods.

Figure \ref{fig:trajectory} illustrates the geometry of conebeam CT with circular trajectories.
Here, consider a plane that is parallel to the $z$-axis and intersects the source trajectory at two locations $\ab(\lambda^-)$ and $\ab(\lambda^+)$. Let $\Pc$
denote such a plane of interest. Then, the main goal of the factorization methods \cite{dennerlein2008factorization,yu2006region}
is to convert the 3-D reconstruction problem to a successive 2-D problems on the planes of interest.
Specifically, on a plane of interest $\Pc$, we define  virtual chord lines  and virtual source locations.
The virtual chord line coordinate system is defined by the virtual chord line direction $\eb$, the $z$-axis $\eb_z$,
 and their perpendicular axis $\eb^\perp$.
The virtual sources $\ab_v(\lambda^+),\ab_v(\lambda^-)$ can be simply computed as 
$$\ab_v(\lambda)=\ab(\lambda)+z\eb_z . $$
Then,  Cartesian coordinate $\x$ is now converted to a new coordinate $(t,s,z)$
such that
\begin{equation}\label{eq:xyz}
\x =  t\eb+s\eb^\perp+ z\eb_z~.
\end{equation}
On a given plane of interest $\Pc$  with a fixed $s$,  with a slight abuse of notation,
 the object density and the DBP data can be represented by the following 2-D functions
$$f(t,z) := f(\x(t,z)),\quad g(t,z):=g(\x(t,z)) \ .$$

\begin{figure}[!t] 	
\centerline{\includegraphics[width=0.7\linewidth]{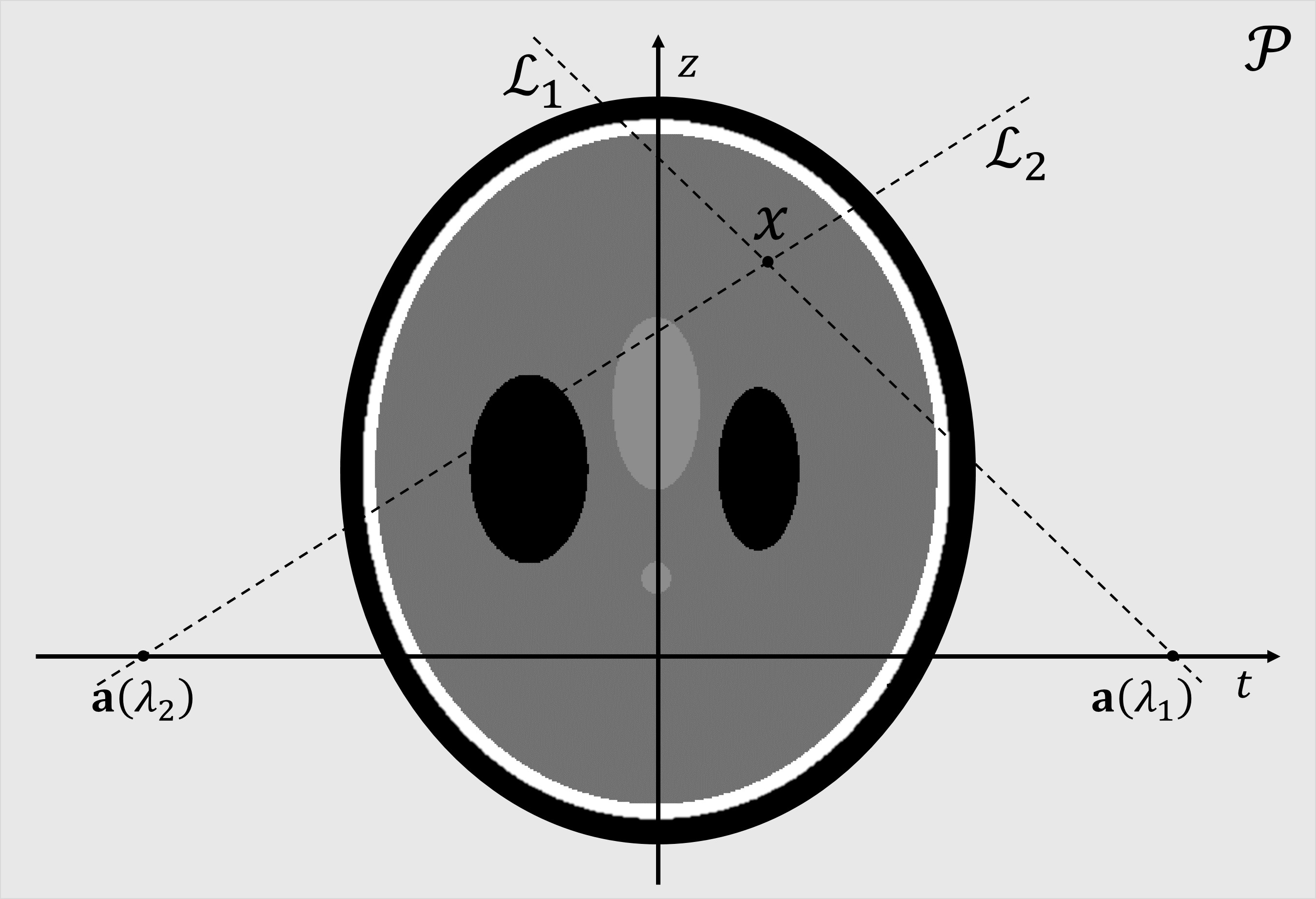}}
\caption{Source geometry and filtering directions on the plane of interests. }
\vspace*{-0.5cm}
\label{fig:aliasing}
\end{figure}

Since the first and the second term in  $\sigma(\x,\omegab,\lambda^\pm)$ in \eqref{eq:sigma}
corresponds to the Hilbert transform along {$\db^-(\x)$ and $\db^+(\x)$} direction, respectively,
%as shown in Fig.~\ref{fig:aliasing}.
%Based on this observation,  one of the most important contributions 
the authors  in \cite{dennerlein2008factorization}  showed that  \eqref{eq:gfourier} can be simply represented as
\begin{eqnarray}
g(t,z)
= \pi \int_{-\infty}^\infty h_H(t-\tau)\left( f(\tau, z_1(\tau)) + f(\tau, z_2(\tau)\right) d\tau \label{eq:gtz},
\end{eqnarray}
where $h_H(t)$ denotes the Hilbert transform, and
$z_1(\tau)$ and $z_2(\tau)$ refer to the coordinate along $\Lc_1$ and $\Lc_2$ lines as shown
in Fig.~\ref{fig:aliasing}.
This implies that the conebeam reconstruction problem can be addressed by solving a deconvolution
problem in the DBP domain.

\section{Main Contribution}
\label{sec:theory}

\subsection{Encoder-Decoder CNNs for Deconvolution on $\Pc$}

%Note that the relationship between the DBP data $g(t,z)$ and the unknown image $f(t,z)$  in \eqref{eq:gtz} is basically a convolution
%relationship, so
Although \eqref{eq:gtz} suggests that  a deconvolution algorithm can recover the original image $f(t,z)$ from $g(t,z)$ on each plane of interest $\Pc$, 
 there are two  technical difficulties.
First, the grid for the unknown image is not a standard cartesian grid due to the $t$ dependent $z_1(t)$ and $z_2(t)$ coordinates.
In fact,  $f(t,z_1(t))$ and $f(t,z_2(t))$ can be regarded as deformed images of $f(t,z)$ on new coordinate systems.
Therefore, the convolution filter is spatially varying, so standard deconvolution methods do not work.
Second, the deconvolution problem is highly ill-posed, since each DBP data $g(t,z)$ has contribution from two deformed images,
$f(t,z_1(t))$ and $f(t,z_2(t))$. 
Accordingly,  the authors in  \cite{dennerlein2008factorization} proposed a  regularized matrix inversion approach
 after the  discretization of the intergration with respect to  the original fixed cartesian grid for the unknown images.
This makes the deconvolution algorithm by Dennerlein et al \cite{dennerlein2008factorization} 
computationally expensive and sensitive to many hyper-parameters.
{On the other hand,  once the neural network is trained,
%neural network approaches have fewer number of hyper-parameters  that can be learned
%from the training phase.
% Moreover, 
 the inference stage of the neural network does not require any computationally expensive optimization method, which makes the algorithm
significantly faster.}

% However, storing such matrix requires significant amount of memory.

In general,  a deconvolution algorithm for \eqref{eq:gtz} can be represented as a mapping:
\begin{eqnarray}
 \Tc: g \mapsto f,\quad g \in \Xc, f \in \Yc
\end{eqnarray}
where  $\Xc$ denotes the input space where DBP data $g(t,z)$ lives, and
$\Yc$ refers to the space that the image $f(t,z)$ belongs.
%
%and  $\Tc$ is  an inverse mapping from the DBP  $g\in \Xc$ to the unknown image $f \in \Yc$.
For the case of Tikhonov regularization, $\Tc$ has a closed form expression; but for general
regularization functions such as $l_1$ or total variation (TV), the inverse mapping $\Tc$ is generally
nonlinear, and should be found using computationally expensive iterative methods.
%However, the iterative methods are usually computational expensive. 
A quick remedy to reduce the run-time computational complexity
would be precalculating nonlinear mapping $\Tc$.  Unfortunately, as the mapping $\Tc$ depends on the input, storing
$\Tc$ for all inputs will  require huge amount of memory and is not even feasible.

In this regard,  an encoder-decoder CNNs (E-D CNN) using ReLU nonlinearities provides an ingenious way of  addressing this issue.
Specifically, in our recent theoretical work \cite{ye2019cnn}, %which is also briefly summarized in Appendix for self-containment,
 we showed that an encoder-decoder CNN with ReLU nonlinearity  is  a piecewise linear approximation
 of a nonlinear mapping, which is composed of large number of locally  linear mappings. % for each region of the input space $\Xc$. 
More specifically, the input space $\Xc$
is partitioned into non-overlapping regions where input for each region shares the common linear representation.
Then,  the switching to the corresponding linear representation for each input 
can be done instantaneously based on the ReLU activation
patterns.
We further showed that although the piecewise linear property of CNN may provide a limitation to approximate arbitrary nonlinear functions, it also provides good architectural prior known as inductive bias, which results in inherent regularization effects. As will be shown in experimental results, we found that this inductive basis works favorable for conebeam artifact removal by providing more generalization power.  

Another unique aspect of deep neural network is that these exponentially many linear representations can
be derived from a small set of filter sets thanks to the combinatorial nature of the ReLU nonlinearities. 
Specifically, the network training is to estimate the filter set $\Thetab$:
\begin{eqnarray}
\min_{\Thetab} \sum_{i=1}^N\|f^{(i)} - \Tc_\Thetab g^{(i)} \|_2^2,
\end{eqnarray}
where $\{(f^{(i)},g^{(i)})\}_{i=1}^N$ denotes the training data set composed of ground-truth image and  DBP image,
and $$\Tc_\Thetab: g\mapsto f,\quad g\in \Xc, f \in \Yc$$ refers to the inverse mapping parameterized by $\Thetab$.
Usually,  filter parameters $\Thetab$ require much smaller memory and space. % than saving $\Tc$ for each input $g$.
On the other hand, the number of associated linear representations increases exponentially with the network depth, width, and skipped connection.
This exponential expressivity and the input adaptability make the neural network powerful in approximating
nonlinear function with relatively small number of parameters.
%In addition to the former claim, in this revision we have specifically mentioned that the CNN is indeed a piecewise linear approximation of the nonlinear mapping, which may have additional regularization effects.  

\subsection{Spectral Blending}

\begin{figure}[!t]
    \centering
	\begin{minipage}[b]{0.6\linewidth}
		 \centerline{\includegraphics[width=\linewidth]{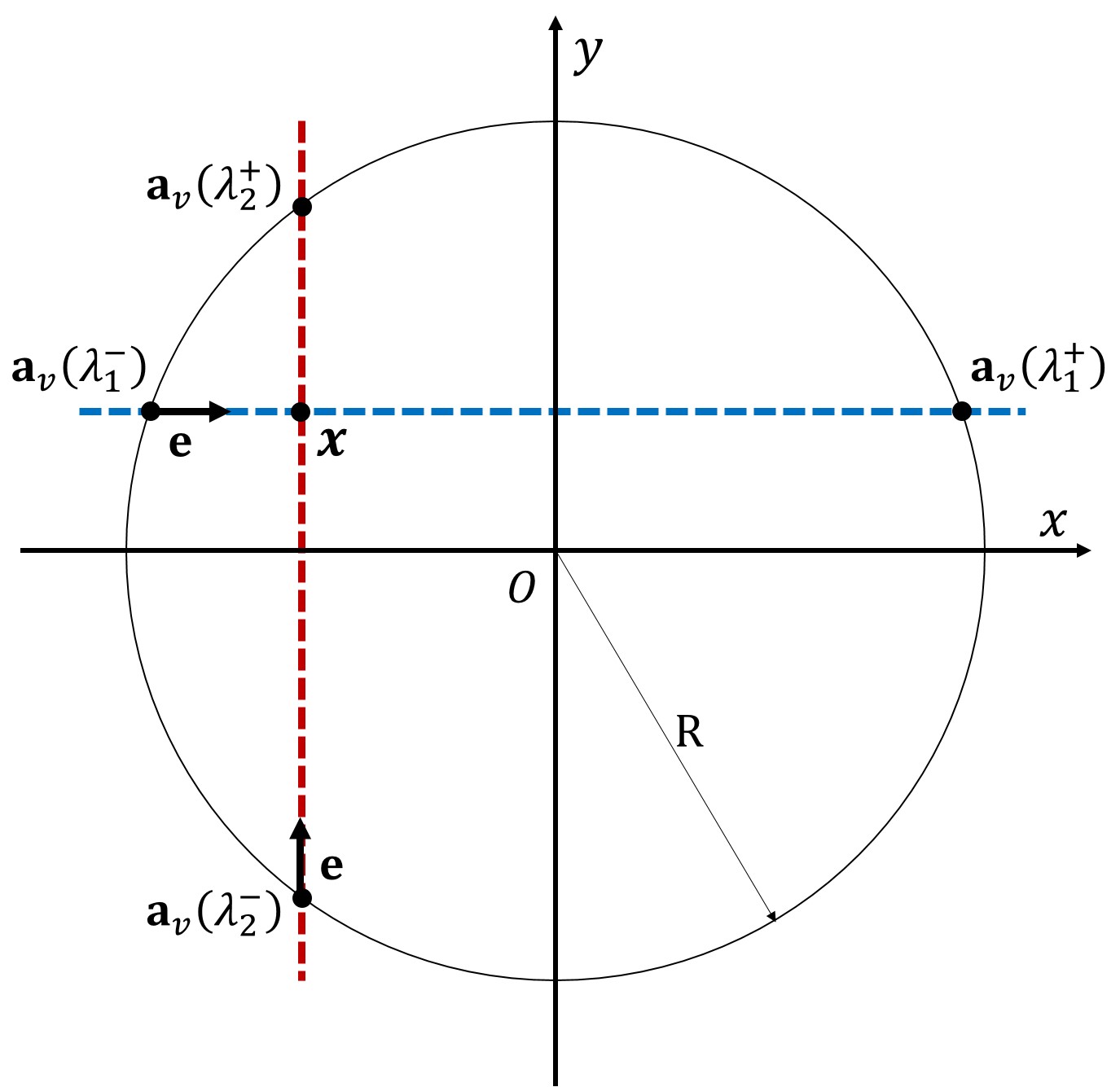}}
		\centerline{(a)}\medskip
	\end{minipage}
    \begin{minipage}[b]{\linewidth}
	\begin{minipage}[b]{0.3\linewidth}
		\centerline{\includegraphics[width=\linewidth]{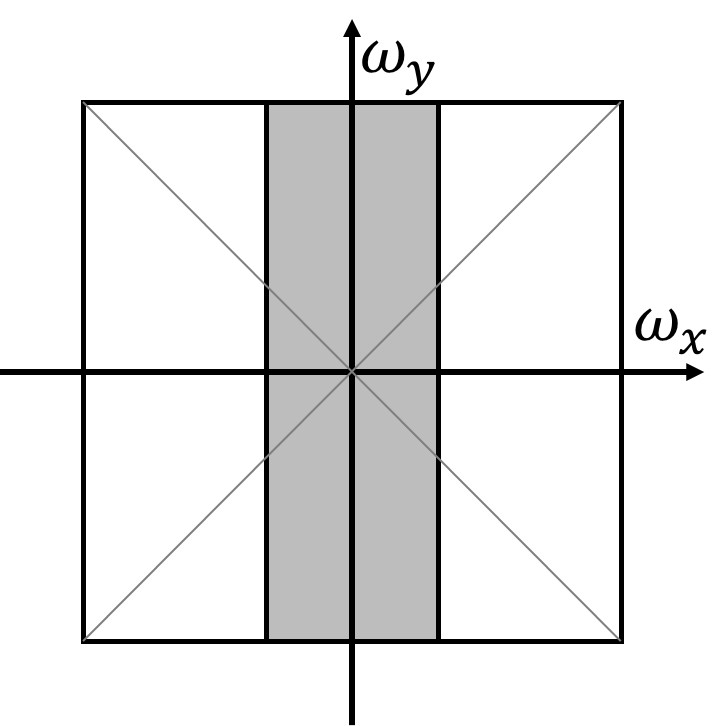}}
		\centerline{(b)}\medskip
	\end{minipage}
	\begin{minipage}[b]{0.3\linewidth}
		\centerline{\includegraphics[width=\linewidth]{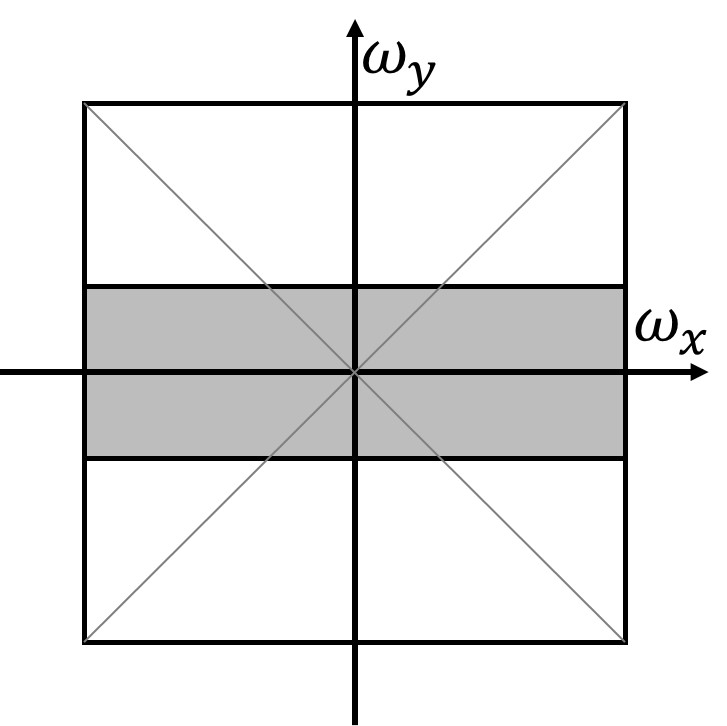}}
		\centerline{(c)}\medskip
    \end{minipage}
    \begin{minipage}[b]{0.3\linewidth}
    	\centerline{\includegraphics[width=\linewidth]{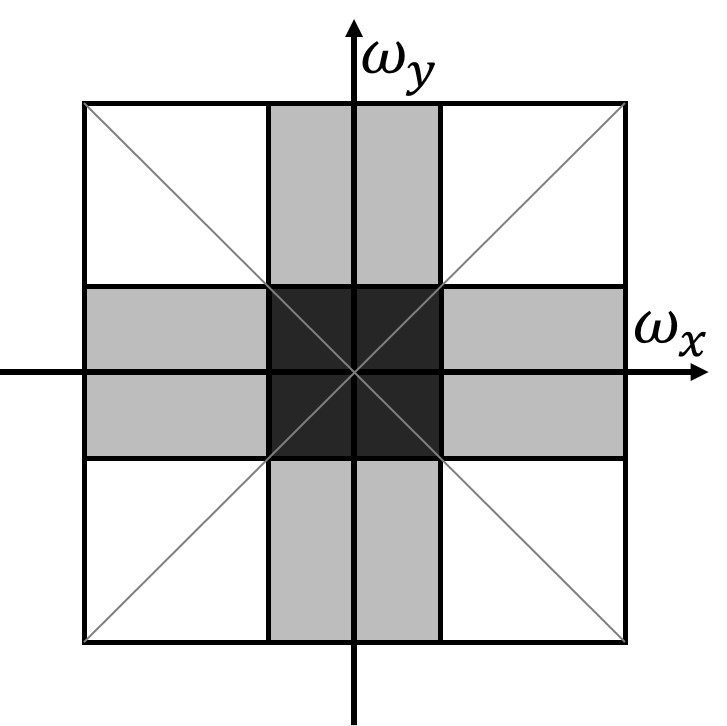}}
    	\centerline{(d)}\medskip
    \end{minipage}
	\end{minipage}
    \caption{The missing frequency region. (a) The top view of a point $\x=(x,y,z)$ and an source trajectory. The {blue} line indicates the coronal directional plane of interest,  and the
    {red} line denotes the sagittal directional plane of interest.
 (b)(c) The missing frequency region for coronal and sagittal directional processing, respectively.     (d) The dark region denotes a common missing frequency region, respectively. 
    }
    \label{fig:missing_frequency_region}
\end{figure}

Although the factorization method leads to 2-D deep learning approach for each plane of interest,
 Theorem~2.2 in our previous work \cite{lee2015interior} showed that 
  the missing frequency region depends on the direction of the plane of interest.
  However,  we also showed in our prior work \cite{lee2015interior} that at any point $\x$, if the missing frequency regions for different filtering directions are appropriately combined, we can minimize the artifacts from the missing frequency regions.  Accordingly, this section describes this in detail.

%These phenomenon can be analyzed more quantitatively using \eqref{eq:f_g_hbt}.
Suppose that we are interested in recovering a voxel at $\x=(x,y,z)$ from the DBP data.  Figure \ref{fig:missing_frequency_region}(a) illustrates
the top view of a point $\x$ and a source trajectory.
 %from the actual source locations $\ab(\lambda),\lambda\in[\lambda_1^+,\lambda_1^-]$. 
 In  Figure \ref{fig:missing_frequency_region}(a), $\ab_v(\lambda)$ denotes the corresponding virtual source location.
 Now, consider the image reconstruction along the coronal direction where the plane of interest $\Pc$ is aligned in the horizontal direction
 (for example, between $\ab_v(\lambda_1^-)$ and $\ab_v(\lambda_1^+)$).
According to the spectral analysis in \cite{lee2015interior}, the resulting DBP data 
 has missing frequency regions along sagittal direction as illustrated in Figure \ref{fig:missing_frequency_region}(b).
 Therefore, any deconvolution algorithm using the coronal directional DBP data 
 may have noise boosting along the sagittal direction due to the ill-posedness in the missing frequency regions.
Similarly, for the DBP data along sagittal direction  (for example, between $\ab_v(\lambda_2^-)$ and $\ab_v(\lambda_2^+)$), 
 the corresponding missing frequency region is given by Figure \ref{fig:missing_frequency_region}(c),
 which results in the noise-boosting after the decovolution.
 This leads to an important question: which directional plane we should choose?
 
 For the case of full scan conebeam CT, there are redundancy in the 2-D factorization of the 3-D data, which can be exploited to answer the question.  More specifically,  we can perform reconstruction for both coronal and sagittal directions, and combine them.
Figure \ref{fig:missing_frequency_region}(d) is the resulting missing frequency regions by combining the reconstruction results from two directions.
In particular, the common missing frequency region is a dark square centered at the origin, which is the intersection of the two missing frequency regions
in coronal and sagittal processing.
This implies that except for the common frequency regions, the missing spectral
components in coronal directional processing can be compensated by the results from the sagittal processing, and vice versa.

\begin{figure*}[!bt] 	
\centerline{\includegraphics[width=0.8\linewidth]{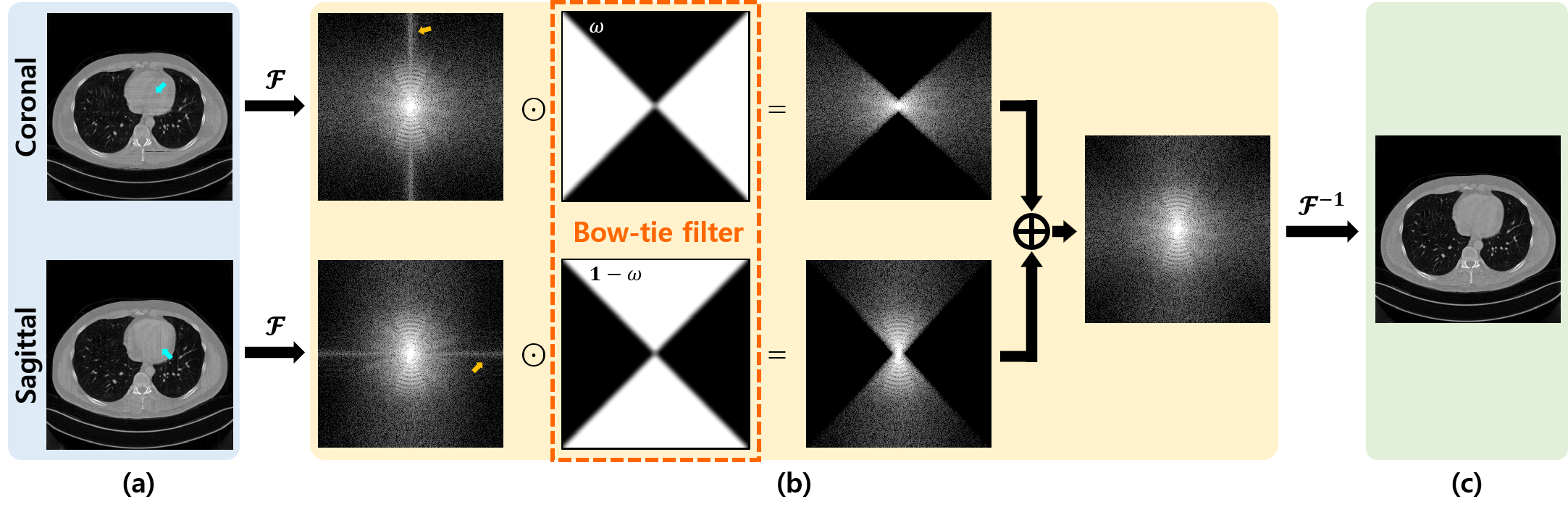}}
\caption{Flowchart for the spectral blending. (a) shows axial images from the coronal and sagittal directional  processed 3-D reconstruction by neural networks.
 (b) illustrates spectral blending, and (c) shows reconstruction result from spectral blending.  }
%\vspace*{-0.5cm}
\label{fig:spectral}
\end{figure*}

Fig. \ref{fig:spectral} shows a flowchart for the spectral blending that was proposed in \cite{lee2015interior} to  achieve this synergistic combination. In particular, the axial images $f^{cor}$ and $f^{sag}$ from coronal and sagittal
reconstructed 3-D volumes are first converted to the spectral domain using the 2-D Fourier transform.
Due to the ill-posedness along the missing frequency regions,
% the deep neural network approaches can fill in the missing frequency regions in Fig.~\ref{fig:missing_frequency_region}(b)(c), 
%this also introduces  artifacts that 
%are described by
streak pattern artifacts, as indicated by the yellow arrows in Fig. \ref{fig:spectral}(b),  are usually visible  in  the Fourier domain along the missing frequency region. 
Then, we apply the bow-type spectral weighting to suppress the signal in the missing frequency regions
and combined them together. This process can be mathematically represented by
\begin{eqnarray*}
	f^{com} = \mathcal{F}^{-1} \left\{ w \odot  \Fc\{f^{cor}\} + (1-w) \odot \Fc\{f^{sag}\}\right\}.
\label{eq:spectral}		 
\end{eqnarray*}
where $\mathcal{F}$ and  $\mathcal{F}^{-1}$ denote the Fourier and inverse Fourier transforms, respectively;
 $w$  and $1-w$ are the bow-tie spectral masks as shown in Fig.~\ref{fig:spectral} (b), and $\odot$ denotes the element-wise multiplication.

%\begin{figure*}[!hbt] 	

\section{Method}\label{sec:method}

\subsection{Numerical Experiments}
%\subsubsection{Supervised learning}

The proposed neural network is formulated in a supervised learning framework, so we need many matched
input and output pairs. To deal with the data issues, the proposed neural network is basically trained using
 simulated noiseless conebeam projection data from ground-truth 3-D volume data.
 Specifically, the conebeam projection data and its processed version (such as DBP data) are used
as the neural network input, and the ground-truth 3-D volume data are used as the label data for supervised training.

\subsubsection{Data Set}

More specifically, to train the neural network using simulated projection data,
simulated projection data from 
ten subject data sets from  American Association of Physicists in Medicine (AAPM) Low-Dose CT Grand Challenge were used.
The data set was selected from the abdominal scans that include liver regions.
The $x-y$ size of images is $512 \times 512$ and the $z$ size ranges from 400 to 600, but varies from the patient to the patient.
The voxel size is 1 mm$^3$. 
 Out of ten patient, eight patient data were used as training sets, one patient data was used as validation set, and the other patient data was used for test set.  To investigate the dependency on the training data size, we also performed separate neural network training by reducing
 the number of training sets.

\begin{figure}[!t] 	
\centerline{\includegraphics[width=1\linewidth]{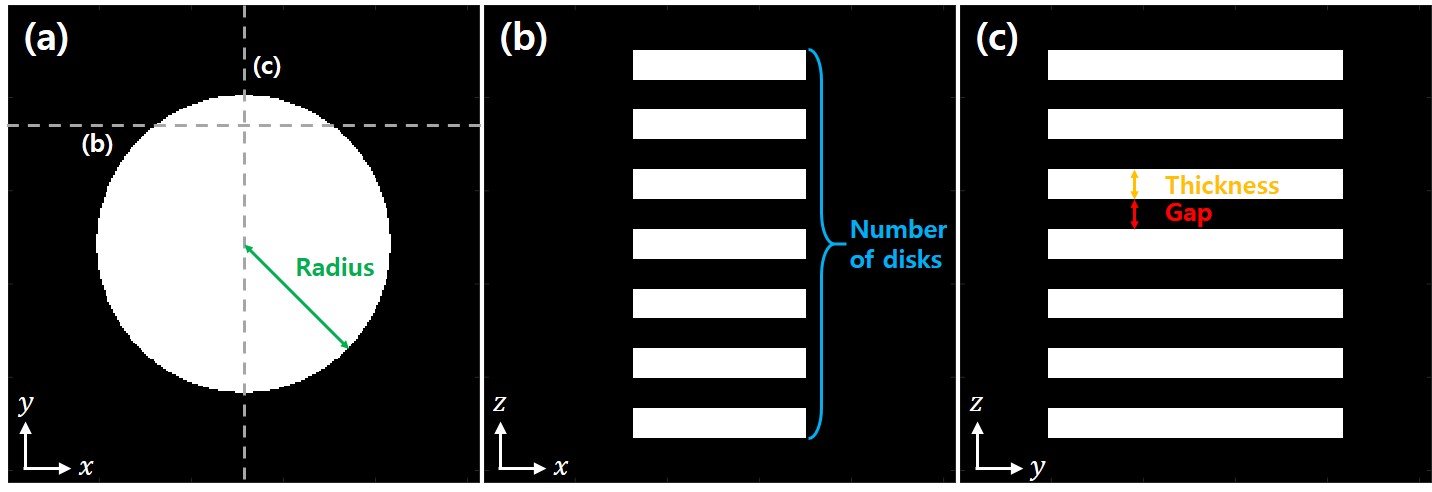}}
\caption{Disk phantom. (a) shows the disk image on the axial plane. (b) and (c) illustrates cut-views from coronal and sagittal planes, respectively.}
%\vspace*{-0.5cm}
\label{fig:phantom}
\end{figure}

 Additionally, we also conducted the experiments
using the
numerical phantoms in  Fig. \ref{fig:phantom}, whose {CT number are either  5,108  Hounsfield Unit (HU) or -1,000 HU}. %is binary, i.e. 0 and 1.
Specifically, Fig. \ref{fig:phantom} shows a typical example of numerical phantom composed of disks with the same thickness and the spacing
between the adjacent disks.
The radius of disk is 80 mm.
By changing the size, radius, the number, thickness, and 
 spacing as shown in Table~\ref{tbl:phantom}, we created various phantoms.
 As we know the ground-truth values, these data set were used
  to investigate the network performance in a more quantitative way.

\begin{table}[!thb]
\centering
\caption{Specification of disk phantom parameters. (a)(b)(c) were used for fine-tuning,
and (d) was used for test. }
\label{tbl:phantom}
\resizebox{0.35\textwidth}{!}{
	\begin{tabular}{c|c|c|c|c}
	\hline
			\ &  (a)  & (b)  & (c) & (d)  \\ \hline\hline
			\ Thickness [mm] & 10  & 14 &  20 &16 \\
			\ Spacing [mm] & 10  & 14 &  20  & 16\\
			\ Number of disks & 9  & 7 &  5  & 7\\
	\hline
	\end{tabular}
		}
\end{table}

\subsubsection{Conebeam Projection Data Generation}

From the volume data set from AAPM and the numerical phantoms,i
conebeam CT sinogram data are numerically generated using numerical cone-beam projection. {No noise is added in this simulated data generation.}
For the numerical projector, the following geometric parameters were used.
The number of detectors was assumed  to be $1440 \times 1440$ elements with a pitch of 1 mm$^2$. The number of views was set to 1200. The distance from source to origin (DSO) was 500 mm, and the distance from source to detector (DSD) was 1000 mm. The maximum cone angle was set to   $35.8 \degree$, and the
projection data were generated accordingly.
%For the image reconstruction
Since our neural network is trained on the plane of interest along the coronal
and sagittal slices,  the training dataset also consists of coronal and sagittal slices. This corresponds to $8192~(512 \times 8 \times 2)$ and $1024~(512 \times 1 \times 2)$ coronal/sagittal slices for training and validation, respectively.  %The size of test image is $512 \times 512 \times 486$. 

% These phantom will be used either for tr
% Here, the same the thickness and the spacing are used for the same factor. 
 %Four phantoms are created using 5 parameters:  The size of disk phantom is $256 \times 256 \times 256$ elements with voxel of $1~\rm{mm}^3$. 
 
\subsection{Real Data Experiments}

We also obtained the real projection data from a real head phantom to validate the generalization power of the proposed
method that is trained with the simulation data.
The head phantom data was acquired from a conebeam CT with circular trajectory with X-ray of 140kVp-9mA. 
 The  detector was $1024 \times 1024$ array matrix with detector pitch of
$0.4 \times 0.4$ mm$^2$ and the number of views was 360.
The distance from source to rotation axis is 1700 mm, and the distance from source to detector is 2250 mm.
The cone-angle of the geometry is $10	.32\degree$.
The reconstruction resolution of the real head 
phantom was $512 \times 512 \times 512$ voxels with voxel size $0.703 \times 0.703 \times 0.703$ mm$^3$.

\subsection{Neural Network Training}

\begin{figure}[!t] 	
\centerline{\includegraphics[width=0.9\linewidth]{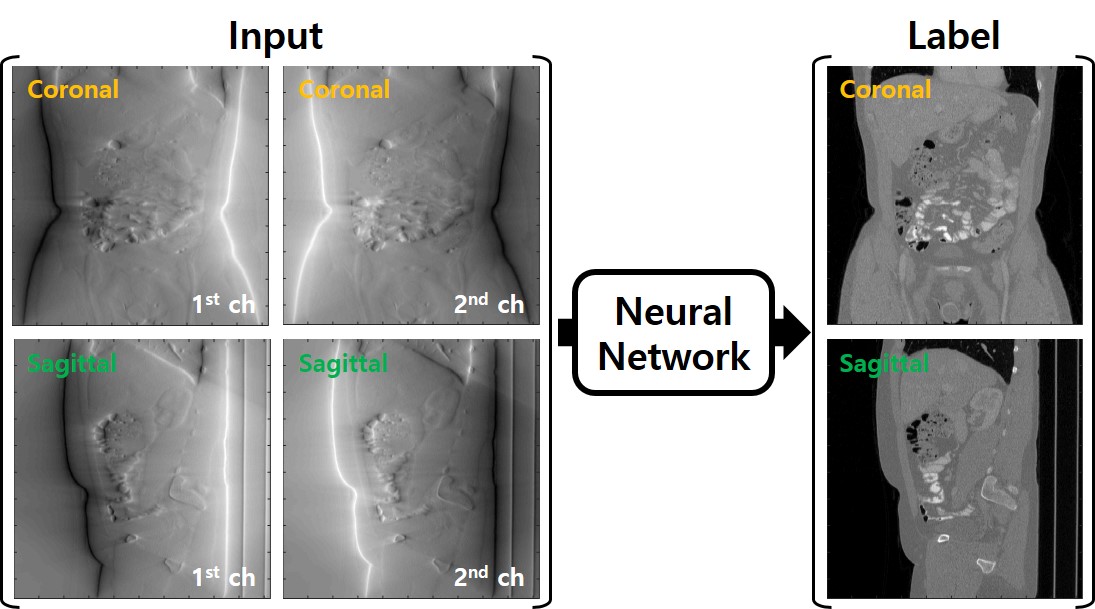}}
\caption{Proposed network input/output for conebeam artifact removal. }
%\vspace*{-0.5cm}
\label{fig:train}
\end{figure}

Fig. \ref{fig:train} illustrates the input and output of the proposed deep learning methods.
Both coronal and sagittal view DBP images are used as inputs,  and the artifact-free coronal and sagittal view images
are used as labels.
For the full scan conebeam CT, the DBP images on the plane of interest $\Pc$ can be obtained using the two complementary source trajectories,
i.e. $\ab(\lambda)$ for $\lambda \in [\lambda^-,\lambda^+]$ and $\lambda \in [0,2\pi]\setminus [\lambda^-,\lambda^+]$, respectively, 
where $\setminus$ denotes the set complement, i.e. $A\setminus B = A \cap B^c$.
In this case, we use both DBP images as inputs as shown in Fig. \ref{fig:train}.
For data augmentation, the whole data set were performed with vertical flipping. {The size of the mini-batch was 4 }%The mini-batch was used as 4 
and the size of input patch is 256 $\times$ 256. Since  convolution operations can be applied for different size images, 
the trained neural network were used for $512\times 512$ images at the inference phase.

\begin{figure}[!t] 	
\centerline{\includegraphics[width=1\linewidth]{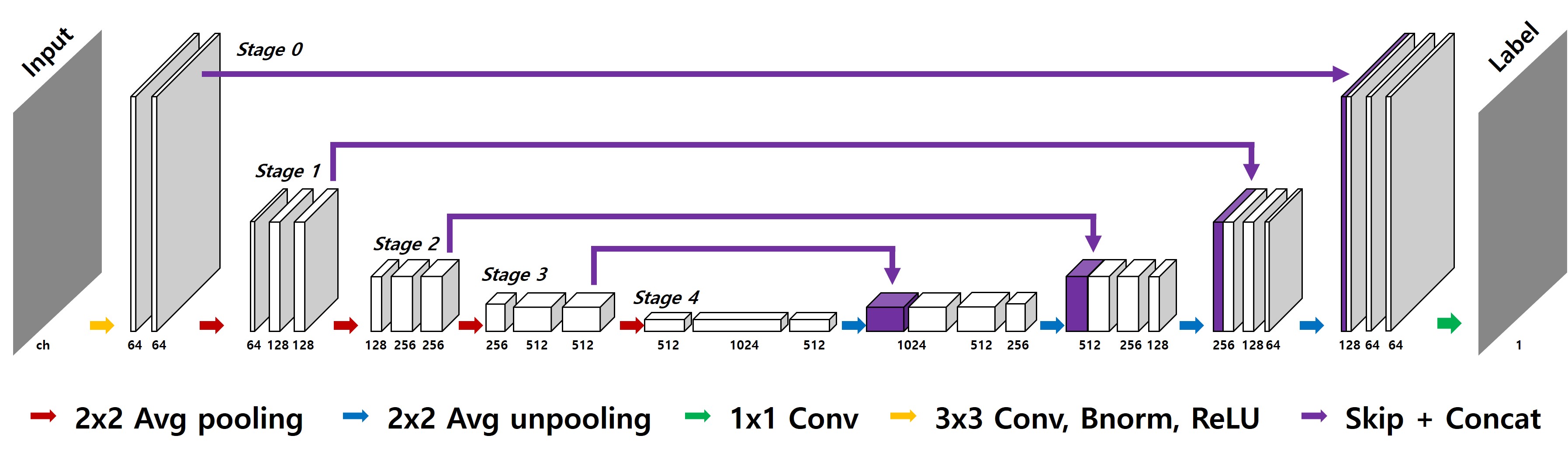}}
\caption{U-Net  backbone for our method and FDK-CNN. }
%\vspace*{-0.5cm}
\label{fig:network}
\end{figure}

The network backbone corresponds to a modified architecture of U-Net \cite{ronneberger2015u} as shown in Fig. \ref{fig:network}.
A yellow arrow in Fig. \ref{fig:network} is the basic operator and consists of $3 \times 3$ convolutions followed by a rectified linear unit (ReLU) and batch normalization.
The yellow arrows between the separate blocks at each stage are not shown for simplicity.
%The architecture consists of a feature extraction part and an feature contraction path.
A red arrow is a $2 \times 2$ average pooling operator and located between the stages.
%Average pooling operator doubles the number of channels and reduces the size of the layers by four.
Average pooling operator reduces the size of the layers by four.
In addition, a blue arrow  is $2 \times 2$ average unpooling operator, reducing the number of channels by half and increasing the size of the layer by four. 
A violet arrow is the skip and concatenation operator. %, connecting directly paired layers in the featuer extraction path and in the feature contraction path.
A green arrow is the simple $1 \times 1$ convolution operator generating final reconstruction image.
The total number of trainable parameters is about 22,000,000.

%\begin{figure}[!b] 	
%\centerline{\includegraphics[width=1\linewidth]{fig/objective}}
%\caption{Convergence plots for the objective function. [Train] : training curve. [Valid] : validation curve. }
%%\vspace*{-0.5cm}
%\label{fig:objective}
%\end{figure}

MatConvNet toolbox (ver.24) \cite{vedaldi2015matconvnet} was used to implement FDK and DBP domain networks in MATLAB R2015a environment. 
Processing units used in this research are Intel Core i7-7700 (3.60 GHz) central processing unit (CPU) and GTX 1080 Ti graphics processing unit (GPU).
The $l_2$ loss was used as the objective function of the training, and the number of epochs for training networks were 300. Stochastic gradient descent (SGD) method was used as an optimizer to train the network. 
The initial learning rate was $10^{-4}$, which gradually dropped to $10^{-5}$ at each epoch. 
The regularization parameter was $10^{-4}$.
%The size of entire input data is 512 $\times$ 512. 
%Training time was about 4 days. 
%Fig. \ref{fig:objective} shows the convergence plot for  each network. The dashed and solid lines represent the objective function of the training and validation phases, respectively. Since the training and validation curves converge closely, we concluded that the networks are well-trained and not over-fitted. 

%Fig.~\ref{fig:phantom} shows the numerical disk phantom structure, and Four phantoms are created using 5 parameters: size, radius, the number of disks, thickness and spacing between adjacent disks. The size of disk phantom is $256 \times 256 \times 256$ elements with voxel of $1~\rm{mm}^3$. The radius of disk is 80 mm, the thickness and the spacing are used for the same factor. The number of disks differs for each phantom and the specification is shown in Table \ref{tbl:phantom}.

Since the numerical phantoms have different pixel distribution from those of the AAPM data set,
the neural  networks should suffer from domain shift. Thus,  fine-tuning using different disk phantoms is necessary to deal with the
domain shift \cite{tzeng2014deep}.
Specifically, the disk phantoms using the parameters in (a)(b)(c) of Table \ref{tbl:phantom} was used for fine-tuning,
and a disk phantom of Table \ref{tbl:phantom}(d) was used at the test  phase.

\subsection{Comparative Experiments}

%\subsubsection{{Reference Algorithms}}

\begin{figure}[!h] 	
\centerline{\includegraphics[width=0.8\linewidth]{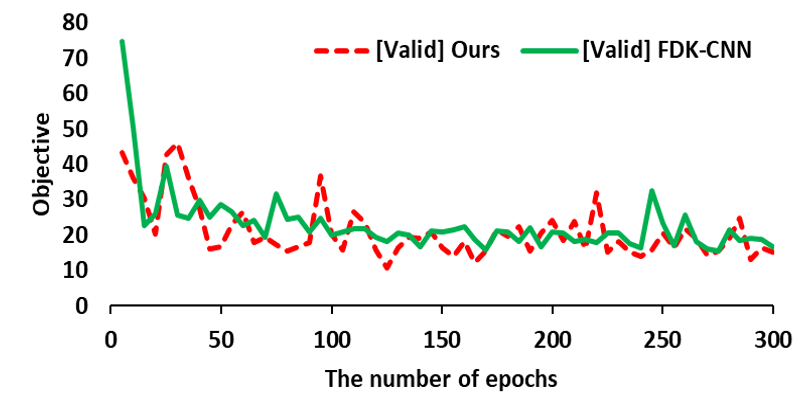}}
\caption{Convergence plots for the objective function on validation set. }
%\vspace*{-0.5cm}
\label{fig:objective_valid}
\end{figure}

As for comparison, we implemented two additional  algorithms:  a deep convolutional  neural network using FDK reconstruction (FDK-CNN), and TV penalized  MBIR algorithm. %econstruction.
The FDK-CNN was trained to  learn the artifact-free images from the reconstructed images of the FDK algorithm using the incomplete projection data due to the cone angle. The input images were FDK images corrupted with the cone-beam artifact, whereas the artifact-free data was used as the ground truth. For a fair comparison, the coronal and sagittal views images are used as the network input  similar to our DBP domain network.
Note that FDK reconstruction combines projections from all full scan angle, so  we cannot obtain the two different FDK images for each processing
direction. This is why only two images along coronal and sagittal views are provided as inputs.
%because the cone-beam artifacts affect the both images at the same time.
The same U-Net architecture was also used for FDK domain neural networks. 
Accordingly, only difference from the proposed DBP domain deep network
 is the input images.
 Fig. \ref{fig:objective_valid} shows the convergence plot for  each network. The plots illustrate the loss curves
calculated after spectral blending at the validation phase. Since the validation curves converge closely, we concluded that the networks are well-trained without over-fitting.

The total variation (TV) penalized reconstruction is also implemented by solving the following optimization problem:
\begin{eqnarray}
	 \min_{\x} \frac{1}{2} \| y - A f \|_2^2 + \lambda\left({\|\nabla_x f\|_1 +\|\nabla_y f\|_1 + \|\nabla_z f\|_1}\right), \nonumber
%	\rm{PSNR} &=& 20 \cdot \log_{10} \left(\frac{NM\|x^*\|_\infty}{\| x- x^*\|_2}\right) \  .
\end{eqnarray}
where $y$ and $f$ are measured sinogram data and its 3-D image, $A$ denotes a system matrix, $\nabla_{x, y, z}$ denotes differentiated operator along $(x, y, z)$-axes, and $\lambda$ is a regularization factor.

\begin{figure*}[!tb] 	
\centerline{\includegraphics[width=0.9\linewidth]{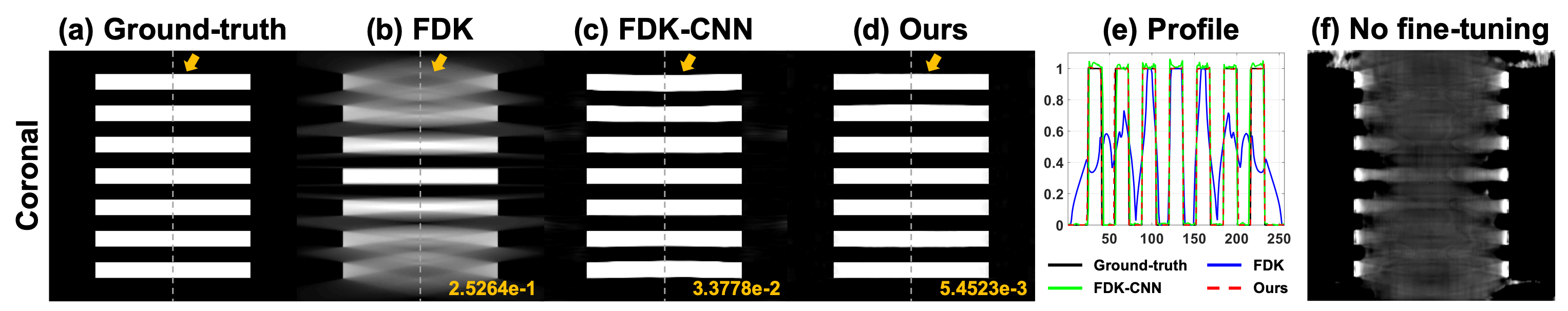}}
\caption{(a) and (b) are the ground-truth image and cone-beam reconstruction by FDK. Disk phantom reconstruction results from (c) FDK-CNN, and (d) proposed method. (e) shows the profiles along the gray line of the images. The number written to the images is the NMSE value. 
%{The values of the black and white part of the images are 0 and 1 respectively. (f) Reconstruction result from the neural network without fine-tuning.}
 The values of the black and white part of the images are 0 and 1 respectively except (f). The window range of (f) is set to [0.48, 0.6] for better visualization.
}
\vspace*{-0.5cm}
\label{fig:result_phantom}
\end{figure*}

%\subsubsection{{{Quantitative Image Metric}}}

For quantitative evaluation, the normalized mean square error (NMSE) value was used, which is defined as
\begin{eqnarray}
	\rm{NMSE} &=& \frac{\sum_{i=1}^{M} \sum_{j=1}^{N} [f^*(i,j) - {f}(i, j)]^2}{\sum_{i=1}^{M}\sum_{j=1}^{N}[f^*(i,j)]^2},
\end{eqnarray}
where $f$ and $f^*$ denote the reconstructed images and ground truth, respectively. $M$ and $N$ are the number of pixel for row and column.
We also use the peak signal to noise ratio (PSNR), which is defined by
\begin{eqnarray}
	\rm{PSNR} &=& 20 \cdot \log_{10} \left(\frac{NM\|f^*\|_\infty}{\| f- f^*\|_2}\right) \  .
\label{eq:psnr}		 
\end{eqnarray}
We also use the structural similarity (SSIM) index  \cite{wang2004image}, defined as
\begin{eqnarray}
	\rm{SSIM} &=& \dfrac{(2\mu_{f}\mu_{f^*}+c_1)(2\sigma_{f f^*}+c_2)}{(\mu_{f}^2+\mu_{f^*}^2+c_1)(\sigma_{f}^2+\sigma_{f^*}^2+c_2)},
\end{eqnarray}
where $\mu_{f}$ is an average of $f$, $\sigma_{f}^2$ is a variance of $f$ and $\sigma_{f f^*}$ is a covariance of $f$ and $f^*$. 
There are two variables to stabilize the division such as $c_1=(k_1L)^2$ and $c_2=(k_2L)^2$.
$L$ is a dynamic range of the pixel intensities. $k_1$ and $k_2$ are constants by default $k_1=0.01$ and $k_2=0.03$.\\

\begin{figure*}[!t] 	
\centerline{\includegraphics[width=0.8\linewidth]{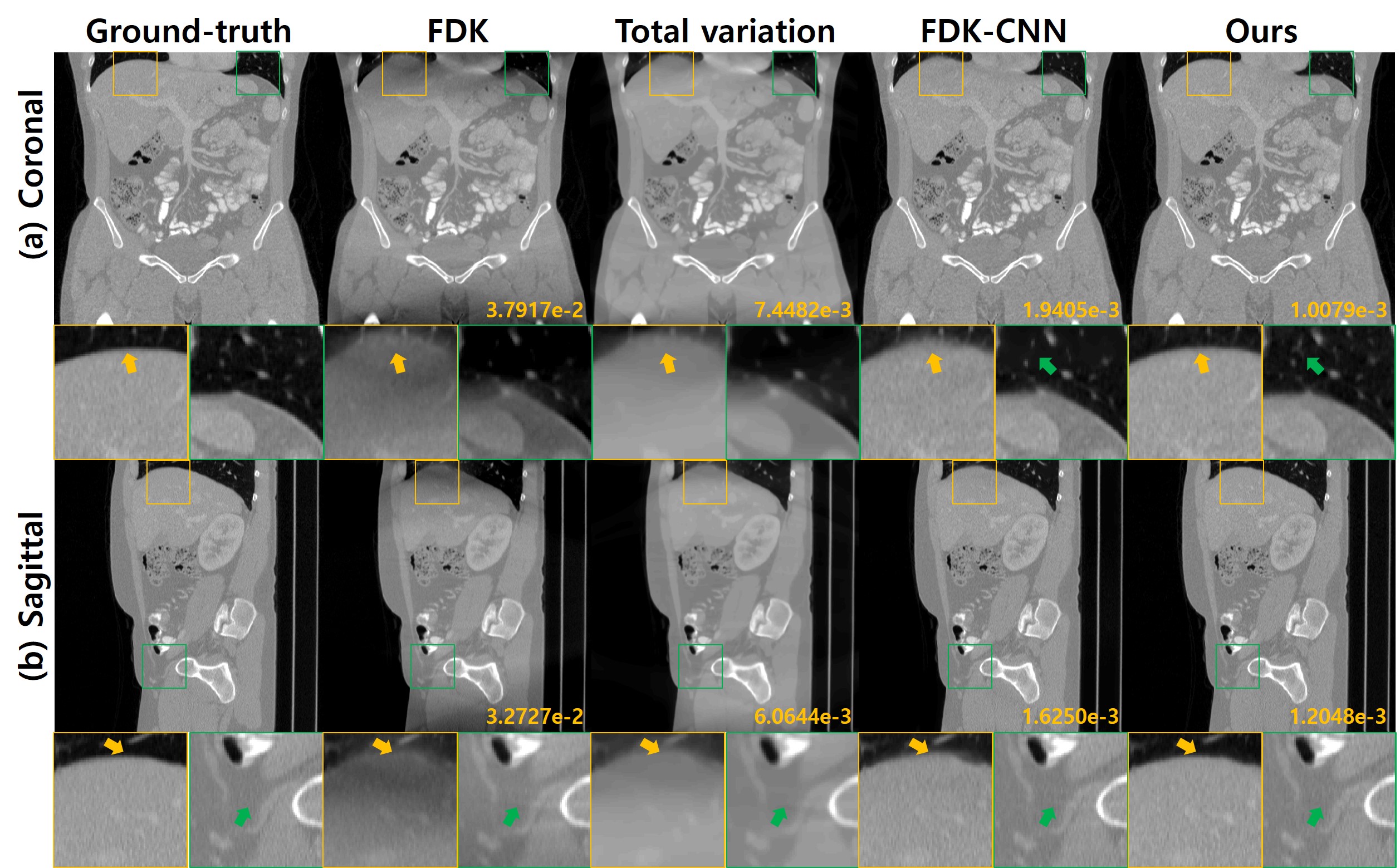}}
\caption{(a) Coronal and (b) sagittal reconstruction results from cone-beam geometry. Yellow and green boxes illustrate the enlarged views of different parts. The number written to the images is the NMSE value. Intensity range of the CT image is (-1000, 800) [HU].}
%\vspace*{-0.5cm}
\label{fig:result_coronal_sagittal}
\end{figure*}

\begin{figure*}[!t] 	
\centerline{\includegraphics[width=0.8\linewidth]{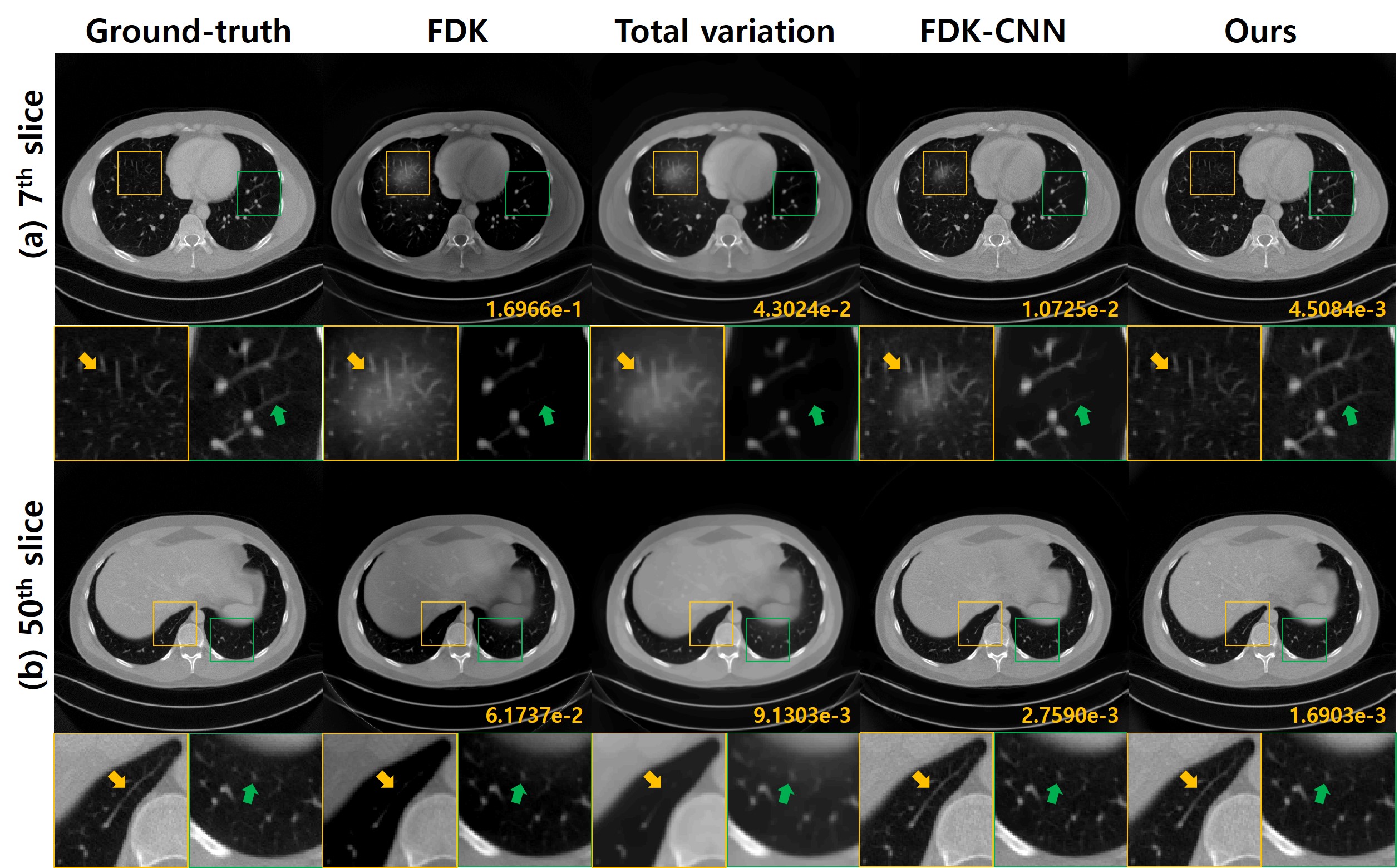}}
\caption{Ground-truth and reconstruction results from FDK, Total variation, FDK-CNN, and proposed method. Yellow and green boxes illustrate the enlarged views of different parts. The number written to the images is the NMSE value. The z-location of (a) is  +236.6mm and (b) is  +193.5mm and the intensity range of the CT image is (-1000, 800)  [HU].}
%\vspace*{-0.5cm}
\label{fig:result_axial}
\end{figure*}

\section{Experimental Results}\label{sec:result}

\subsection{Numerical phantom results}

First, the reconstruction results using numerical phantom is illustrated in Fig.~\ref{fig:result_phantom}. In the FDK reconstruction result in
 Fig.~\ref{fig:result_phantom}(b),  severe cone-beam artifacts were observed especially in the regions far
 from the mid-plane. 
 FDK-CNN  reconstruction results in  Fig.~\ref{fig:result_phantom}(c) can significantly improve the reconstruction
 quality, but  the shape of the disc gradually bends as it moves away from the midplane.
 However, our reconstruction method provided near perfect reconstruction without any  bending  artifacts
 as observed in  Fig.~\ref{fig:result_phantom}(d). 
 Quantitative NMSE value on each figure clearly confirmed that 
 the proposed method significantly outperform other methods.
 
 When FDK-CNN and our neural network trained with AAPM data were directly used for the binary data without fine tuning,
 the results in  Fig.~\ref{fig:result_phantom}(f) shows that the data is not correctly recovered.
 This is as expected since the data distribution of the numerical phantom in Fig.~\ref{fig:result_phantom}(a) is binary with
 piecewise constant regions, which was never seen in AAPM data. Accordingly,
the domain adaptation with
 fine-tuning using binary data was necessary.

%\begin{figure}[!t] 	
%\centerline{\includegraphics[width=1\linewidth]{fig/phantom}}
%\caption{Disk phantom. (a) shows the disk image on the axial plane. (b) and (c) illustrates cut-views from coronal and sagittal planes, respectively.}
%\vspace*{-0.5cm}
%\label{fig:phantom}
%\end{figure}

%\begin{table}[!thb]
%\centering
%\caption{Quantitative comparison for various algorithms. }
%\label{tbl:metric}
%\resizebox{0.48\textwidth}{!}{
%	\begin{tabular}{c|c|c|c|c}
%		\hline
%			\ {Metric}					& {FDK}			& {Total variation}	& FDK-CNN 		& Ours 				\\ \hline\hline
%			\ PSNR [dB]					& {32.0547}		& {34.3922}			& {39.9860}		& {\textbf{42.4859}}	\\
%			\ NMSE ($\times 10^{-3}$)	& {35.728}		& {7.3193}			& {1.7983}		& {\textbf{0.9373}}	\\		
%			\ SSIM						& 0.7767		& 0.7750			& {0.9129}		& {\textbf{0.9684}}	\\ \hline\hline
%			\ Time [sec/slice] 			& - 			& 14.06 			& 0.10			& 0.12 				\\ \hline
%	\end{tabular}
%		}
%\end{table}

\begin{table}[!thb]
\centering
\caption{Quantitative comparison for various algorithms. }
\label{tbl:metric0}
\resizebox{0.48\textwidth}{!}{
	\begin{tabular}{c|c|c|c|c|c}
		\hline
			\ {Metric}					& {FDK}			& {TV}	& FDK-CNN 		& Ours$_{~\rm{half}}$		& Ours 				\\ \hline\hline
			\ PSNR [dB]					& 32.05		& 34.39		& 39.99		& 41.76		& \textbf{42.49}	\\
			\ NMSE ($\times 10^{-3}$)	& 35.73		& 7.32		& 1.80		& 1.26			& \textbf{0.94}	\\		
			\ SSIM						& 0.78		& 0.78		& 0.91	  & 0.97		& \textbf{0.97}	\\ \hline\hline
			\ Time [sec/slice] 			& - 		& 14.06 	& 0.10		& 0.12		& 0.12 				\\ \hline
	\end{tabular}
		}
\end{table}

\subsection{AAPM data results}

Fig. \ref{fig:result_coronal_sagittal} shows the reconstruction images  of AAPM data along the  coronal and sagittal planes. As shown in Fig. \ref{fig:result_coronal_sagittal}, the FDK reconstructions around the midplane are similar to  the ground-truth images, but poor intensity and the cone-beam artifact are visible in the off-midplanes such as top and bottom areas. %Similar to Fig. \ref{fig:result_axial},
Although TV and FDK-CNN methods can improve the image quality, there are still
remaining  conebeam artifacts, and the reconstruction do not preserve the detailed structures and textures. 
On the other hand,  our method clearly removes the artifact, maintains sophisticated structures and textures. 
Fig. \ref{fig:result_axial} shows the axial view of the reconstruction results. The FDK image shows poor intensity and produces cone-beam artifacts due to the higher cone angles. While the TV and FDK-CNN methods compensate for the poor intensity, the cone-beam artifact still remained. In addition, the TV method does not preserve the detail texture, and the FDK-CNN overestimates in Fig. \ref{fig:result_axial}(a) or underestimates in Fig. \ref{fig:result_axial}(b). However, the ours method preserved the detailed texture and clearly removed the cone-beam artifacts as shown in the enlarged images of Fig. \ref{fig:result_axial}.

%Quantitatively, our method has the lowest NMSE values compared to other methods. 
Table \ref{tbl:metric0} shows the quantitative comparisons of average PSNR, NSME, and SSIM and their computation times. Our method produces the best quantitative values in all metrics. 
In addition, we show the quantitative metric when  our network was trained with dataset with only 50\% of training data set. Even though the training set size was halved, our network shows superior result to FDK-CNN in various quantitative meaures. 
The computation time of the our network is slightly slower than that of  FDK-CNN because the input size of our method is twice that of the FDK-CNN, but the neural network methods are much faster than TV method. %In particular, the input of FDK-CNN is single channel, but the our method is two channels.

\subsection{Real Experimental Data}

\begin{figure}[!t] 	
\centerline{\includegraphics[width=0.8\linewidth]{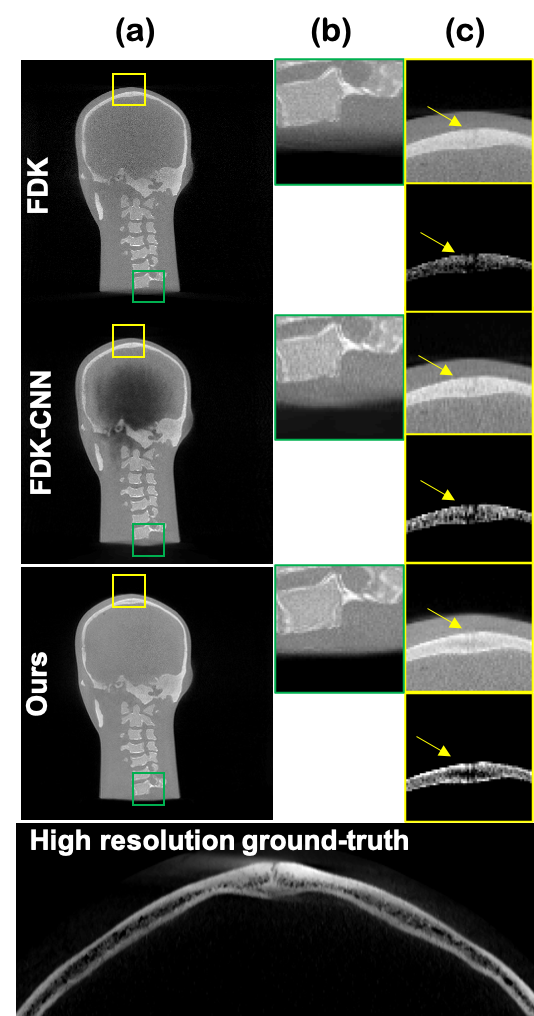}}
\caption{\added{(a) Results of head phantom without fine-tuning the network trained on AAPM datasets. The intensity range of the images is (-1000, 800)[HU].
 Magnified views of (b) green box area, and (c) yellow-box area in two window levels (top: (-1000,800)[HU],   bottom: (300,700)[HU]).  (d) A high resolution CT image is also illustrated as a ground-truth image.}
%{The top and bottom area with conebeam artifacts are magnified.}
}  
%\vspace*{-0.5cm}
\label{fig:head_phantom}
\end{figure}

Finally, we applied the proposed neural network, which was trained using only simulation data, to the real measurement
to investigate its robustness to real system imperfections.
Unlike the experiment for disk phantom, here we did not perform any fine-tuning, since we do not have ground-truth data.
Despite this, the structure is well preserved by our method 
as shown in Fig. \ref{fig:head_phantom}(a). {Aside from conebeam artifact, %Interestingly,
the FDK-CNN produces additional artifacts  inside of head. This issues will be discussed in more detail later. } 

%The real data results confirmed
% that trained network  using the DBP data generalize better than FDK-CNN. 
 Although the conebeam artifact from FDK is not {significant as in the simulation} due to the small cone-angle, 
the {zoomed areas at the top of the skull and the bottom of the the last lower vertebrae in the coronal direction still show conebeam artifact}. 
{Specifically, \added{as shown in  Fig. \ref{fig:head_phantom}(a)(b)}  the tissue boundary at the lower vertebrae is a bit blurry in the FDK reconstruction due to the conebeam artifacts, whereas
the boundary is clear in our method.}
 {Additionally,  \added{ in  Fig. \ref{fig:head_phantom}(a)(c)} the crack \added{and the space between the bones} in the top skull was not visible in FDK and FDK-CNN due to the conebeam artifacts, whereas
\added{they are clearly visible} in the proposed method.} 
%{The axial images of the corresponding slices {at the top of the skull} are also illustrated,in which  the skull boundary was only visible by our method.}
{In order to confirm that this \added{crack and space} are really present,  we  conducted additional experiments by placing the suspected region (the top of the skull) at the mid-plane of a high resolution circular conebeam scanner. The detector array was $654\times 664$ with the pitch of $250\mu m$, and the reconstructed image size is $512\times 512$ with the pixel pitch of $310\mu m$.  As shown in the bottom figure of Fig.~\ref{fig:head_phantom}, the resulting high resolution image clearly shows the \added{crack and space} of the skull which coincides with our reconstruction. The results confirmed that the proposed method produces high resolution image with less conebeam artifacts compared to the standard FDK reconstruction.
}
%
%
%%This again clearly confirmed the robustness and accuracy of the proposed method.
%\\

\section{Discussion}\label{sec:discussion}

\subsection{Effects of  spectral blending}

\begin{figure*}[ht] 	
\centerline{\includegraphics[width=0.8\linewidth]{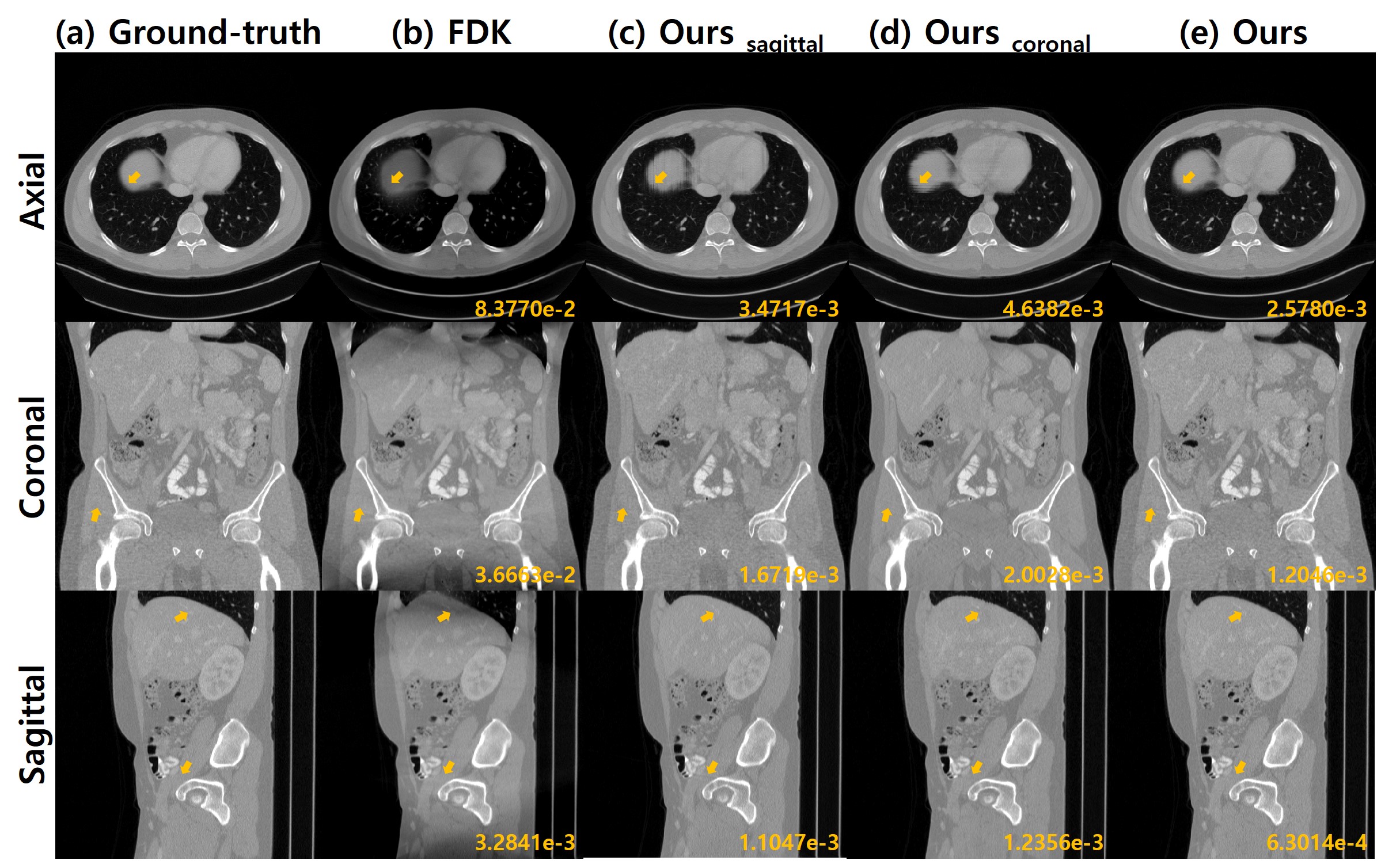}}
\caption{Reconstruction results with and without spectral blending. The number written to the images is the NMSE value. (a) and (b) are the ground-truth image and FDK reconstructed result. (c) is reconstructed into a single network  of sagittal DBP input, but (d) is reconstructed from coronal DBP input. (e) is reconstructed from spectral blending between (c) and (d).  Intensity range of the CT image is (-1000, 800) [HU].}
%\vspace*{-0.5cm}
\label{fig:result_spectral_blending}
\end{figure*}

To demonstrate the importance of spectral blending,  the reconstruction results  by separate processing
along the coronal and the sagittal direction are shown  in Fig. \ref{fig:result_spectral_blending}(c)(d), whereas the results from  the spectral blending  is shown in Fig. \ref{fig:result_spectral_blending}(e). The horizontal and vertical  streaking artifacts patterns are visible in the axial view images
in Fig. \ref{fig:result_spectral_blending}(c)(d), respectively, where the artifact patterns appears along the direction of  plane of interests. 
On the other hand, the spectral blending removes the streaking patterns as shown in  Fig. \ref{fig:result_spectral_blending}(e).
 In the coronal and sagittal view reconstruction results, the results without the spectral blending exhibited remaining streaking patterns, whereas our method
 with spectral blending has successfully removed them.
The PSNR values in Table~\ref{tbl:metric_comp_ch} also confirmed that the spectral blending improves the reconstruction quality.

\begin{table}[!thb]
\centering
\caption{Quantitative comparison to validate the importance of spectral blending. }
\label{tbl:metric_comp_ch}
\resizebox{0.5\textwidth}{!}{
	\begin{tabular}{c|c|c|c}
	\hline
			\ Metric 			&  Ours$_{~\rm{sagittal}}$ & Ours$_{~\rm{coronal}}$ & Ours  \\ \hline\hline
			\ PSNR [dB]  				& 40.88  	& 39.54 				& \textbf{42.49} \\
			\ NMSE ($\times 10^{-3}$)	& 1.30		& 1.73					& \textbf{0.94}	\\ 
			\ SSIM 						& 0.96		& 0.94					& \textbf{0.97}	\\ \hline
	\end{tabular}
		}
\end{table}

\begin{table}[]
\centering
\caption{Quantitative comparison for various experimental conditions} 
\label{tbl:metric}
\resizebox{0.5\textwidth}{!}{
\begin{tabular}{c|c|c|c|c|c}
\hline
\multicolumn{3}{c|}{}                                            & PSNR {[}dB{]}  & NMSE($\times10^{-3}$)  & SSIM          \\ \hline \hline
\multirow{8}{*}{} & \multirow{2}{*}{1} & FDK-CNN & 34.00          & 7.31          & 0.76          \\  
                                         &                    & Ours    & \textbf{38.62} & \textbf{2.41} & \textbf{0.92} \\ \cline{2-6} 
       (a) Number of                                & \multirow{2}{*}{3} & FDK-CNN & 37.18          & 3.30          & 0.90          \\  
                 training                        &                    & Ours    & \textbf{40.43} & \textbf{1.57} & \textbf{0.94} \\ \cline{2-6} 
                       data sets           & \multirow{2}{*}{5} & FDK-CNN & 39.64          & 2.11          & 0.91          \\  
                                         &                    & Ours    & \textbf{41.76} & \textbf{1.21} & \textbf{0.96} \\ \cline{2-6} 
                                         & \multirow{2}{*}{8} & FDK-CNN & 39.99          & 1.80          & 0.91          \\  
                                         &                    & Ours    & \textbf{42.49} & \textbf{0.94} & \textbf{0.97} \\ \hline
%                                     \hline
%\multicolumn{3}{c|}{Metric}                                             & PSNR {[}dB{]}  & NMSE ($\times 10^{-3}$) & SSIM          \\ \hline \hline
\multirow{12}{*}{} & \multirow{3}{*}{16.1$\degree$} & FDK     & 37.28          & 7.83          & 0.89          \\  
                                      &                        & FDK-CNN & 39.11          & 2.15          & 0.92          \\ 
                                      &                        & Ours    & \textbf{39.74} & \textbf{1.39} & \textbf{0.94} \\ \cline{2-6} 
                                      & \multirow{3}{*}{19.8$\degree$} & FDK     & 36.92          & 9.89          & 0.88          \\  
                                      &                        & FDK-CNN & 39.51          & 1.73          & 0.93          \\  
      (b) Cone                                &                        & Ours    & \textbf{40.33} & \textbf{1.22} & \textbf{0.95} \\ \cline{2-6} 
        angle                              & \multirow{3}{*}{25.6$\degree$} & FDK     & 35.92          & 14.45         & 0.87          \\  
                                      &                        & FDK-CNN & 40.23          & 1.41          & 0.94          \\  
                                      &                        & Ours    & \textbf{41.47} & \textbf{0.96} & \textbf{0.97} \\ \cline{2-6} 
                                      & \multirow{3}{*}{35.8$\degree$} & FDK     & 32.05          & 35.73         & 0.78          \\  
                                      &                        & FDK-CNN & 39.99          & 1.80          & 0.91          \\  
                                      &                        & Ours    & \textbf{42.49} & \textbf{0.94} & \textbf{0.97} \\ \hline
 %% noise
\multirow{9}{*}{} & \multirow{3}{*}{15.0dB} & FDK     & 25.90          & 4.68          & 0.55          \\ 
                              &                        & FDK-CNN & 26.50          & 3.95          & 0.66          \\ \ 
                              &                        & Ours    & \textbf{31.29} & \textbf{0.97} & \textbf{0.69} \\ \cline{2-6} 
  (c) Measurement                           & \multirow{3}{*}{16.5dB} & FDK     & 27.42          & 4.13          & 0.62          \\ 
  SNR                            &                        & FDK-CNN & 30.46          & 1.24          & 0.73          \\  
                              &                        & Ours    & \textbf{34.23} & \textbf{0.48} & \textbf{0.80} \\ \cline{2-6} 
                              & \multirow{3}{*}{19.2dB} & FDK     & 30.09          & 3.69          & 0.75          \\ 
                              &                        & FDK-CNN & 36.20          & 0.33          & 0.86          \\  
                              &                        & Ours    & \textbf{38.69} & \textbf{0.18} & \textbf{0.93} \\ \hline

\end{tabular}
}
\end{table}

\subsection{Effects of training data set size}

To investigate the performance degradation with respect to the training data size,
 we also trained our neural network using various training data set size. 
 Specifically, among the 8 patient data set that were originally assigned for the training data set,
we randomly chose 1$\sim$8 patient data set to train the proposed method and the FDK-CNN approach.
  As shown in Table~\ref{tbl:metric}(a), the proposed method outperformed the FDK-CNN across all training data set size. 
  Moreover, the proposed method trained with 3 data set outperformed the FDK-CNN trained with 8 data set.  This again confirms the superior generalization capability of the proposed method.
%Especially, the FDK-CNN hardly learn cone-beam artifact removal on only one volume of training set while ours do well on it. It means training networks on DBP domain result in consistent performance although FDK-CNN has drastically bad result on a small data set.

\subsection{Effects of unmatched cone-angles at the test phase}

Although we trained the neural network using  the projection data from the cone-angle of 35.8$\degree$,
at the test phase the trained neural network was applied for the projection
data from   various cone-angles from 16.7$\degree$ to 35.8$\degree$. This experiment
was  to verify the generalization capability of the proposed method for cone-angle variations.
% as shown in Table R1. The results clearly show that the proposed method provides superior and stable performance compared to the FDK-CNN. 
%We demonstrate our that algorithm is stable for various settings like cone-angle. We change the cone-angle only to $16.07\degree$, $19.80\degree$, $25.64\degree$, $35.80\degree$ to test performance of the proposed method for more realistic settings. 
As shown in Table~\ref{tbl:metric}(b), although the proposed method and FDK-CNN were trained using the projection data from the cone-angle of $35.8\degree$, our method outperforms the FDK-CNN for other cone-angle than FDK-CNN. This again confirms the generalization
capability of the proposed method, which is particular useful for interventional imaging set-up, where the maximum cone-angle
can be controlled by a beam blocking filter. 
{Although in this experiment the neural network was trained only using larger cone angle data,
the results could be further improved by training the network by combining the smaller angle data as training data. }

\begin{figure*}[!t] 	
\centerline{\includegraphics[width=0.8\linewidth]{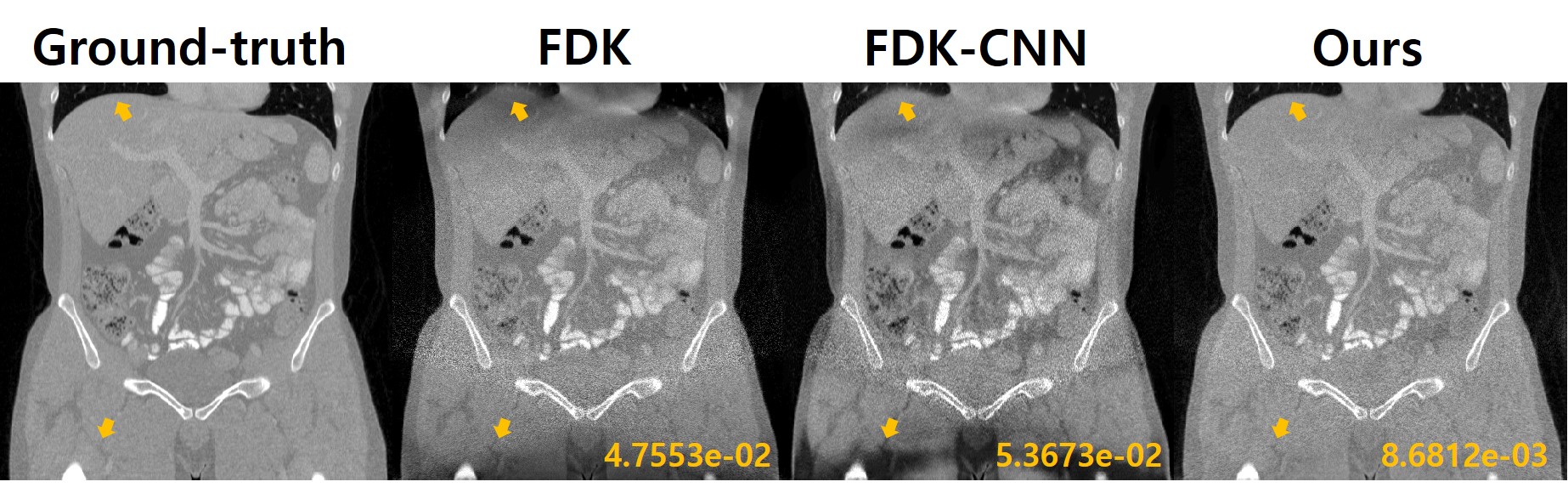}}
\caption{Reconstruction results for noise simulation. The number written to the images is the NMSE value. Initial X-ray intensity is $5 \times 10^5$ {which result in 15.0dB of SNR} and the intensity range of the CT image is (-1000, 800) [HU].}
%\vspace*{-0.5cm}
\label{fig:result_noise}
\end{figure*}

%\begin{figure*}[!t] 	
%\centerline{\includegraphics[width=0.8\linewidth]{fig/result_noise_graph}}
%\caption{{Qualitative comparison for noise simulation. }}
%%\vspace*{-0.5cm}
%\label{fig:result_noise_graph}
%\end{figure*}

\subsection{Noise robustness}
Recall that our network was trained without adding noises.
To demonstrate the noise robustness of our method, 
we added noises to the projection data  at the test phase %when initial X-ray intensity is $5 \times 10^{5}$, $1 \times 10^{6}$, $5 \times 10^{6}$,
using the Poisson noise models. The noise level was selected to result in 15.0, 16.5, and 19.2dB in SNR.
Five realization of noises were generated  and the resulting reconstruction results were averaged. % metrics in Fig. \ref{fig:result_noise_graph} are averaged. 
We compare the result from our network to FDK-CNN. Fig. \ref{fig:result_noise}  shows typical
reconstruction results by various algorithms. In particular,
the FDK-CNN produces more noticeable artifacts compared to the
FDK reconstruction. % It means that the training the network on image domain is vulnerable to noise. 
On the other hand, the proposed network trained on the DBP domain was very robust for different noise levels.
The quantitative comparison in Table~\ref{tbl:metric}(c) also confirmed that the proposed method outperforms other methods.
This again confirms the generalization power of the proposed method.
%The network on DBP domain faithfully is learnt to remove cone-beam artifact while maintaining noise pattern. This is because the network is learnt using given information, not additional information like noise as shown in \cite{lee2019contrast}

\subsection{Generalization to real data}

%We have also applied our neural network trained with the simulation data to the true data set. 
%Unlike the experiment for disk phantom, we test the real head phantom data for the trained network without any fine-tuning. As shown in Fig. , the structure is well preserved by our method, while it is not reconstructed at all by 
Recall that FDK-CNN produces severe artifacts in Fig.~\ref{fig:head_phantom}.
Now, the simulation  results in  Fig. \ref{fig:result_noise} clearly explains the origin of the artifacts from FDK-CNN  in Fig.~\ref{fig:head_phantom}. Specifically, the real projection measurements contain noises that are not considered during the
training. Therefore, when a trained FDK-CNN neural network, which does not generalize well, was applied to the real projection measurement,
this limited generalization power may have produced the artifacts. On the other hand, the proposed neural network trained using the DBP data 
is less sensitive to the noises as discussed before, so that when it is applied to the real measurement we did not observe any
artifacts. 
 It proves that  the network trained on the DBP domain generalize well so that it works even on the data with unseen anatomical region and noise patterns. 

%\subsection{{Unsupervised training}}

\subsection{Limitations of the current study}
While the paper provides promising results for conebeam artifact removal, the current study has several limitations.
First, as shown in disk phantom experiments in Fig.~\ref{fig:result_phantom}(e), the domain shift for unseen anatomical
regions that has different
CT number distribution could be a potential issue in practice, which deserves further investigation.   

Although the generating training dataset in simulation setup is relatively simple, and in real clinical setting it is very difficult to obtain the matched reference data set. To address this problem, we need an unsupervised training scheme, where the artifact-free images are obtained from other person or scans from complete trajectory (such as spiral scan in medical CT).  In fact,  our previous work \cite{kang2019cycle} demonstrated that unsupervised learning using CycleGAN architecture  provides excellent reconstruction results for the low-dose CT applications and
we believe that similar unsupervised learning approach may be possible for conebeam artifact removal.
This is beyond the scope of the current paper, and will be reported in a separate paper.
 
% and we believe that DBP domain deep learning is an essential building block for both supervised and unsupervised learning setup.  
Finally, in real conebeam acquisition scenario using circular trajectory, there are additional sources of artifacts, among which
scatter artifacts from the lack of scatter grid  is one of the most dominating artifacts.  In the current study, the simulated
projection data does not assume any scatter events, which may lead to another difficulty in applying the proposed method
in clinical settings. Therefore, more systematic study using unmatched data set is necessary to validate the proposed network
architecture for real clinical environment.

\section{Conclusions}\label{sec:conclusion}

In this paper, we developed  a novel DBP domain deep learning approach for conebeam artifacts removal.
Inspired by the existing factorization  approach that converts 3-D problem into a successive 2-D deconvolution along the plane of interest,
our neural network is designed as a 2-D neural network for each plane of interest and was trained to learn the mapping between DBP data and the artifact-free images.
Furthermore, spectral blending technique, which mixes the spectral components
of the coronal and sagittal reconstruction, was employed
 to mitigate the missing frequency occurred along the  plane of interest direction. Experimental results showed that the our method
 generalizes much better than CNN approach using FDK and 
  is quantitatively and qualitatively superior to the existing methods,  despite significantly reduced computational runtime.
%
%
%neural network trained with dual channel DBP inputs, and demonstrated the superiority of dual DBP input with regard to reconstruction of the cone-beam artifacts. The DBP domain neural network learned the inverse Hilbert transform with projection onto the range space of the artifact free images, and the proposed dual channel network has better generalization and specificity than FDK domain neural network. 

\section{Acknowledgement}

The authors would like to thanks Dr. Cynthia MaCollough, the Mayo Clinic, the American Association of Physicists in Medicine (AAPM), and grant EB01705 and EB01785 from the National Institute of Biomedical Imaging and Bioengineering for providing the Low-Dose CT Grand Challenge data set.
{The authors would like to thank Prof. Seungryoung Cho and Mr. Jaehong Hwang to providing the real skull phantom experimental data.}

\bibliographystyle{IEEEtran}
\bibliography{ref}

\end{document}